\documentclass[structabstract]{aa}
\usepackage{amsmath}
\usepackage{txfonts}
\usepackage{graphicx}
\usepackage{amsbsy}
\usepackage{bm}
\usepackage{tablefootnote}
\usepackage{color}
\usepackage{natbib}

\DeclareMathOperator{\sgn}{sgn}

\usepackage{natbib}
\bibpunct{(}{)}{;}{a}{}{,}

\begin{document}

\title{Chasing the peak: optimal statistics for weak shear analyses}
\author{Merijn Smit \inst{1}\thanks{\email{msmit@strw.leidenuniv.nl}} \and Konrad Kuijken \inst{1}}
\institute{Leiden Observatory, Leiden University, PO Box 9513, 2300RA Leiden, The Netherlands}
\date{Received 20 June 2017 / Accepted 9 October 2017}

\abstract
{Weak gravitational lensing analyses are fundamentally limited by the intrinsic distribution of galaxy shapes. It is well known that this distribution of galaxy ellipticity is non-Gaussian, and the traditional estimation methods, explicitly or implicitly assuming Gaussianity, are not necessarily optimal.}
{We aim to explore alternative statistics for samples of ellipticity measurements. An optimal estimator needs to be asymptotically unbiased, efficient, and robust in retaining these properties for various possible sample distributions. We take the non-linear mapping of gravitational shear and the effect of noise into account. We then discuss how the distribution of individual galaxy shapes in the observed field of view can be modeled by fitting Fourier modes to the shear pattern directly. This allows scientific analyses using statistical information of the whole field of view, instead of locally sparse and poorly constrained estimates.}
{We simulated samples of galaxy ellipticities, using both theoretical distributions and data for ellipticities and noise. We determined the possible bias $\Delta e$, the efficiency $\eta$ and the robustness of the least absolute deviations, the biweight, and the convex hull peeling estimators, compared to the canonical weighted mean. Using these statistics for regression, we have shown the applicability of direct Fourier mode fitting.}
{We find an improved performance of all estimators, when iteratively reducing the residuals after de-shearing the ellipticity samples by the estimated shear, which removes the asymmetry in the ellipticity distributions. We show that these estimators are then unbiased in the absence of noise, and decrease noise bias by more than $\sim 30\%$. Our results show that the convex hull peeling estimator distribution is skewed, but still centered around the underlying shear, and its bias least affected by noise. We find the least absolute deviations estimator to be the most efficient estimator in almost all cases, except in the Gaussian case, where it's still competitive ($0.83<\eta <5.1$) and therefore robust. These results hold when fitting Fourier modes, where amplitudes of variation in ellipticity are determined to the order of $10^{-3}$.}
{The peak of the ellipticity distribution is a direct tracer of the underlying shear and unaffected by noise, and we have shown that estimators that are sensitive to a central cusp perform more efficiently, potentially reducing uncertainties by more than $50\%$ and significantly decreasing noise bias. These results become increasingly important, as survey sizes increase and systematic issues in shape measurements decrease.}


\maketitle


\section{Introduction}

Since the first gravitational shear detections \citep{Tyson_1990}, the statistical analysis of weak gravitational lensing effects has become recognized as a competitive cosmological tool. With the advent of precision cosmology, meaningful interpretations of statistical agreement or tension between various models and datasets become increasingly important.

Weak gravitational lensing produces slight magnification and distortion effects by bending the paths of light rays. Although analyses of the former have produced important scientific results \citep[e.g.,][]{Mag_Hildebrandt_2009,Mag_Waerbeke_2010} and it has in fact been demonstrated that combined analyses can give better constraints \citep[][]{Mag_Hildebrandt_2011, Mag_Ford_2012}, most scientific information has come from the analysis of weak shear distortions. To access that information, one has to be able to (1) measure the shapes of lensed background sources accurately, (2) understand the intrinsic distribution of these shapes and the effects of shear and noise on statistical inference, and (3) obtain the statistical power to probe the subtle perturbations of this distribution by weak shear.

For the first part, a multitude of shape measurement methods have been explored, among which are foremost methods based on surface brightness moments \citep[e.g.,][]{KSB_1995,Rhodes_2000} and model fitting methods \citep[e.g.,][]{KK_1999,BJ02,HS03,Refregier_2003,KK_2006,Lensfit_Miller07,Lensfit_Kitching08}, with various alternative or combined approaches \citep{Bernstein_2014,Herbonnet_2016,Zhang_2015}.

Community-driven projects for optimal and robust shape estimates \citep{STEP_2006,STEP2_2007,GREAT08_2010,GREAT10_2012,GREAT3_2015} have led to a further decrease in measurement variances and a better understanding of remaining systematic effects and biases \citep[e.g.,][]{Voigt_2010,Bernstein_2010,Noise_Kacprzak12,Noise_Melchior12,Noise_Refregier12}.

For the last part, the last two and a half decades have also known dramatic improvements in statistical power. Surveys that are finished, ongoing, and planned such as COSMOS\footnote{http://cosmos.astro.caltech.edu/} \citep{COSMOS_2007}, CFHTLenS\footnote{http://www.cfhtlens.org/} \citep{CFHTLenS_Heymans12}, RCSLenS\footnote{http://www.rcslens.org/} \citep{RCSLenS_2016}, KiDS\footnote{http://kids.strw.leidenuniv.nl/} \citep{KIDS_2013}, DES\footnote{http://www.darkenergysurvey.org/} \citep{DES_2016}, LSST\footnote{https://www.lsst.org/} \citep{LSST_2008}, Euclid\footnote{http://www.euclid-ec.org/} \citep{Euclid_2011}  steadily increase in size (sky coverage and depth) and imaging quality, including a significant improvement in understanding and correcting for systematic effects \citep[e.g.,][ for CFHTLenS]{CFHTLenS_Rowe12,CFHTLenS_Heymans12}.

This increasing statistical power is necessary to overcome the inference limit set by the intrinsic galaxy shape distribution, known as shape noise. Unlike many forms of noise, such as measurement uncertainties that are often dominated by Poisson processes, there is no reason that the ellipticities of background galaxies follow a Gaussian distribution. In fact, studies of galaxy morphologies \citep{LML92,Rodriguez_intsh} suggest that late type galaxies may exhibit a roughly uniform axis ration distribution.

\begin{figure}[h]
\centering
\resizebox{\hsize}{!}{
\includegraphics{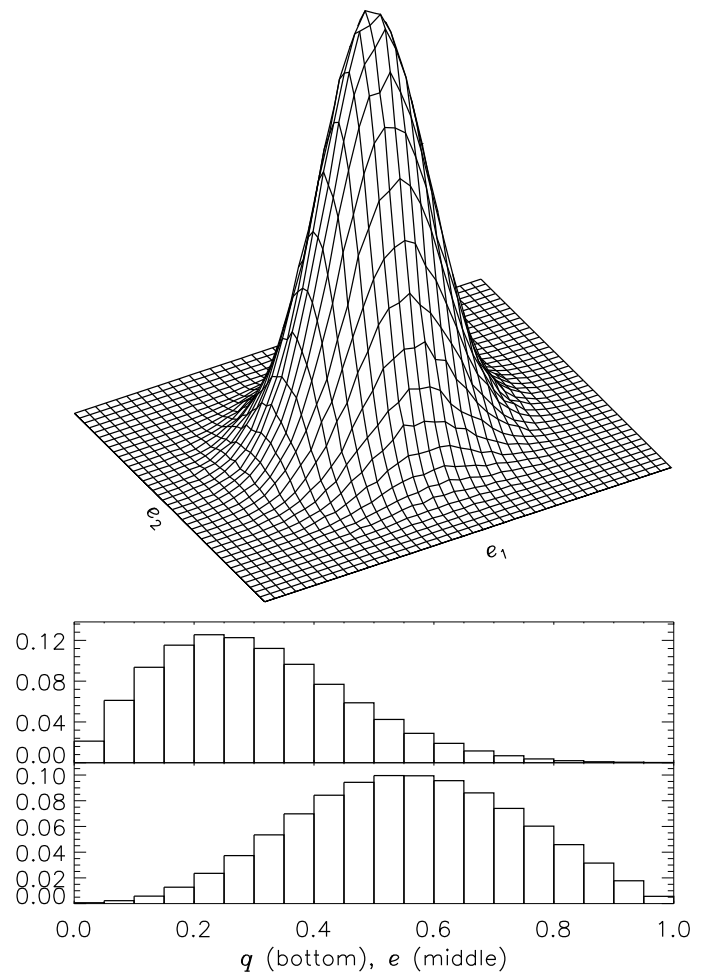}
}
\caption{Gaussian ellipticity distribution and corresponding axis ratio distribution. Top: a 2D histogram of ellipticities. Middle: histogram of the absolute ellipticity $|e|$. Bottom: histogram of the ellipse axis ratio $q$.}
\label{dist_norm}
\end{figure}

\begin{figure}[h]
\centering
\resizebox{\hsize}{!}{
\includegraphics{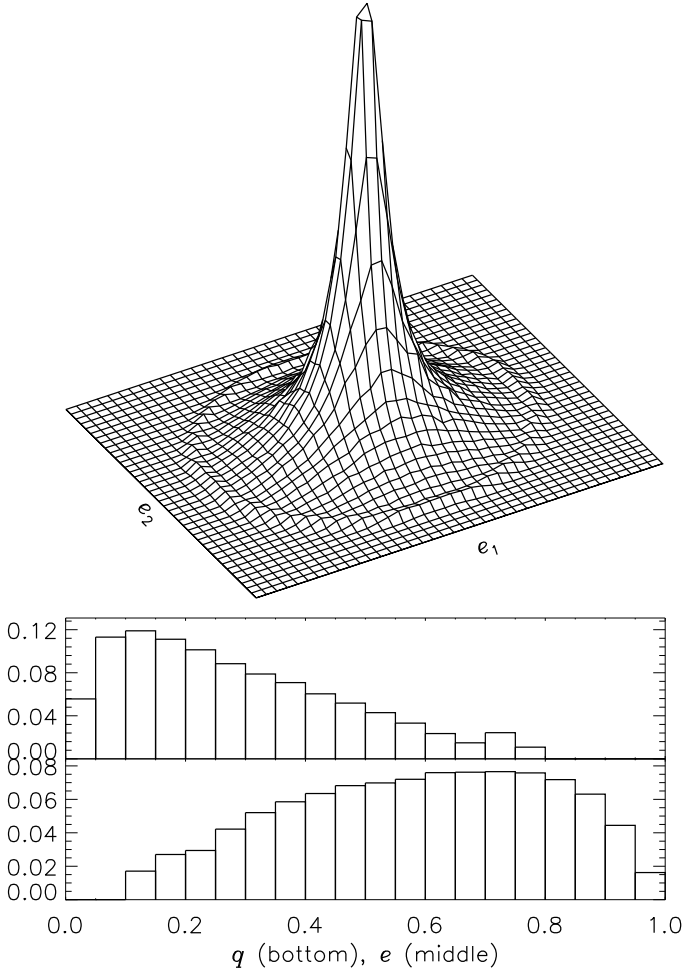}
}
\caption{Ellipticity and axis ratio distributions of the CFHTLenS catalog. Top: a 2D histogram of ellipticities. We note that the ring-like feature at $e\approx 0.8$ is due to noisy outliers forced to a maximum $e$ by the shape measurement pipeline, but see also Figure \ref{fig:dist_stark}. Middle: histogram of the absolute ellipticity $|e|$. Bottom: histogram of the ellipse axis ratio $q$.}
\label{dist_cfhtls}
\end{figure}

This departure from Gaussianity is clearly demonstrated in Section \ref{sec: wl}, when comparing the shape distribution of the CFHTLenS shape measurements catalog \citep[][Figure \ref{dist_cfhtls}]{CFHTLenS_Heymans12} to a simulated Gaussian distribution (Figure \ref{dist_norm}). This implies that commonly used Gaussian estimators, such as the (weighted) mean estimate of the central peak of the distribution or the variance for its width, are not necessarily optimal for the inference of the underlying gravitational shear.

For example, if the tails of the ellipticity distribution decline more slowly than the Gaussian $\exp\left(-x^2\right)$, then more elliptical galaxies contribute more shape noise. There have been many weighting and clipping schemes suggested to minimize biases and uncertainties in weak shear inference \citep{Bonnet_1995, Waerbeke_2000, BJ02}. Alternative approaches include distribution symmetrization \citep{Zhang_2016}, or using ensembles of galaxies in Bayesian analyses or nulling techniques \citep{Bernstein_2014, Herbonnet_2016}, so that the step of individual shape measurement before inference of the underlying shear is bypassed.

In this article, we explore an alternative approach by reviewing statistical estimators that are more suited to a distribution with a pronounced central cusp and slowly declining tails. Estimator optimality would include a low or vanishing estimator bias and a high accuracy by a low spread in estimates. These aspects should be robust for various possible distributions, as samples of background galaxies are comprised of different populations.

We then highlight the use of these estimators in fitting the shear pattern in the field of view with Fourier modes (Fourier Mode Fitting, FMF). This approach provides an alternative to smoothed gridding and locally sparse and therefore poorly constrained estimates. It provides statistical information constrained by the whole field of view, and incorporates fluctuations in background number densities and estimated measurement uncertainties automatically. For subsequent scientific analyses, the Fourier model allows for relatively straightforward, analytic approach to fundamental quantities, such as a power spectrum or mass density reconstruction.

We note that we focus on the statistical inference from samples of measured shapes, for various possible intrinsic shape distributions, that is, the propagation of shape noise. This is a single but fundamental step in improving the accuracy and fidelity of weak lensing analyses. We do not perform a subsequent cosmological analysis, which would require addressing other well-known sources of bias and systematic effects. These include for example selection and detection biases \citep[e.g.,][]{HS04,CFHTLenS_Miller13,DES_2016_cat} among others on the instrumental and computational side. Other sources include physical effects that affects the interpretation of the measured signal, such as the effects of baryons, or the redshift distribution and intrinsic alignments of lensed background galaxies background sources. The shear signal we recover in this paper would represent a combined signal, which would then need to be interpreted.

The remainder of this paper is organized as follows. We will briefly review the necessary definitions of galaxy shapes and the weak lensing formalism in Sect. 2, referring the reader to excellent reviews such as \citet{Review_Bartelmann_2001,Review_Schneider_2006,Review_Hoekstra_2008}, for more in-depth approaches. We review the necessary statistical framework in Sect. 3, where we discuss galaxy shape distributions and statistical estimators, including definitions for efficiency and bias, before expanding on FMF. In Sect. 4 we describe the various possible simulations and data, and analysis methods. In Sect. 5 we discuss the results and the scientific implications. Section 6 gives a summary of our conclusions.

\section{Weak lensing} \label{sec: wl}

Gravitational lensing is the effect of curved space-time on the paths of light rays from distant sources to the observer as they pass through the gravitational potential of foreground structures. This geometrical effect leads to a displacement of point sources on the projected plane of the sky. The differential effect on images $I(x,y)$ of extended sources leads to magnification and distortion effects, know as the convergence $\kappa$ and the shear $\gamma =\gamma_1+ i\gamma_2$, directly related to the surface mass density. This is commonly described as a coordinate transformation \begin{equation}
\label{eq:lens}
\left( \begin{array}{c} x' \\ y' \end{array} \right) =\left( \begin{array}{cc} 1-\kappa-\gamma_1 & -\gamma_2 \\ -\gamma_2 & 1-\kappa+\gamma_1 \end{array} \right) \left( \begin{array}{c} x \\ y \end{array} \right) \,,
\end{equation} resulting in the lensed image $I(x',y')$.

Weak lensing magnification analyses \citep[e.g.,][]{Mag_Hildebrandt_2009,Mag_Waerbeke_2010} require the intrinsic (distribution of) source sizes or magnitudes. In weak shear analyses, the focus lies on the net distortion or reduced shear $g = g_1+ig_2 \equiv (\gamma_1 +i\gamma_2) / (1 - \kappa)$: \begin{equation}
\label{eq:lens_re}
\left( \begin{array}{c} x' \\ y' \end{array} \right) = \left( 1-\kappa \right) \left( \begin{array}{cc} 1-g_1 & -g_2 \\ -g_2 & 1+g_1 \end{array} \right) \left( \begin{array}{c} x \\ y \end{array} \right) \,,
\end{equation} where the transformation is written as a multiplication of $( 1-\kappa )$ (which leads to the magnification) and a traceless distortion matrix describing the alignment of lensed sources in the foreground potential.

The distortion effect of weak lensing shear on images of background galaxies depends on their intrinsic shape distribution. While galaxies often have complex morphologies, it is adequate to describe images by their quadrupole brightness moments or their ellipticities, and the respective response to weak shear distortions.

A common definition of the shape of an image with elliptical isophotes is the ellipticity $e=e_1+ie_2$, defined as the reduced shear needed to create this image from an image with circular isophotes \citep{BJ02,KK_2006}. This gives an axis ratio $q=\frac{b}{a}$ as \begin{equation}
\label{eq:axis}
q=\frac{1-|e|}{1+|e|} \quad \Leftrightarrow \quad |e|=\frac{1-q}{1+q}=\frac{a-b}{a+b} \,,
\end{equation} and position angle $\theta$ via \begin{equation}
\label{eq:posang}
e=|e|\left( \cos{2\theta}+i\sin{2\theta}\right) .
\end{equation}

As an example, we compare a Gaussian $(e_1,e_2)$ distribution to the distribution observed in the CFHTLenS shape measurement catalog in Figures \ref{dist_norm} and \ref{dist_cfhtls}.

This complex notation gives a most straightforward formulation of the resulting ellipticity $\tilde{e}$, after transforming an image with ellipticity $e$ by a distortion $g$, by \citet{SS3_97}

\begin{equation}
\label{eq:SS3_97}
\tilde{e} = \frac{e+g}{1+g^*e} \quad \mathrm{for} \quad |g|\le 1 \,,
\end{equation}with $g^*$ the complex conjugate of $g$.

\begin{figure}[h]
\centering
\resizebox{\hsize}{!}{
\includegraphics[angle=90]{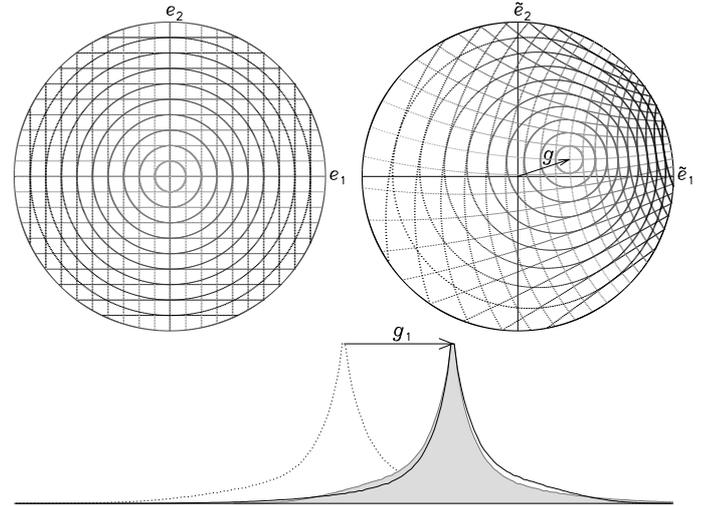}
}
\caption{Top: the non-linear mapping of ellipticities (with $|e|\le 1$) by an exaggerated gravitational shear of $g=0.33+0.11i$. Bottom: the asymmetry introduced in the ellipticity distribution, highlighted for the $e_1$-component.}
\label{fig:mapping}
\end{figure}

The non-linear effect of gravitational shear on the ellipticity parameters is shown in the top panel of Figure \ref{fig:mapping}. Through statistical estimation, we can attempt to infer from an ensemble of galaxy shapes the underlying shear, if we assume the intrinsic ellipticity distribution $P(e)$ to be centered around zero ellipticity. In other words, one assumes no preferred direction on the sky.

This non-linear response to weak shear distortions gives rise to the asymmetry in the observed ellipticity distribution, as shown in the bottom panel of Figure \ref{fig:mapping}. The shifted central peak of the distribution is unaffected by this non-linearity and therefore a direct tracer of $g$.

The canonical approach is a weighted mean $\mu$, where weighting schemes attempt to minimize systematic effects from noise, size and brightness. As observed by \citet{SS3_97}, the expectation value $\left<\tilde{e} \right>$ does not depend on $P(e)$ in the absence of noise. The mean of an ensemble of measured ellipticities is then an asymptotically unbiased estimator for the underlying shear $g$.

In the presence of noise, however, these estimations suffer from unavoidable biases in the estimated shear \citep{Noise_Melchior12,Noise_Kacprzak12}. Furthermore, the variance of an estimator such as the mean, or more generally, the scale of the estimator distribution, does depend on the intrinsic ellipticity distribution $P(e)$. Informally put, the smaller the estimator variance, the more `trustworthy' the estimates and the more efficient the estimator. A more efficient estimator reduces the uncertainties in and therefore the error bars or confidence intervals of parameter estimates.

The smearing of the sheared distribution by noise affects central value estimations, but the peak location itself is still an unbiased tracer of the shear.

\section{Statistical framework}

In this section, we discuss various estimators, after reflecting upon estimator properties, such as bias, efficiency, and robustness, and their interpretation. We then propose ways to apply this to fitting individual Fourier modes to a shear field.

\subsection{Bias, efficiency and robustness} \label{bias_eff_rob}

We will use the term bias, or $\Delta e$, when referring to the difference between the central value of an estimator, such as the expected value or mean $\left<\hat e \right>$, and the population parameter $e$. We will use the term residuals, or $r_i=e_i-\hat e$, when talking about the differences between one sample estimate and the elements of that sample, that is, the individual measurements $e_i=e_{i,1}+ie_{i,2}$.

We note that we write $r_i=e_i-\hat e$ for simplicity throughout this paper, but we employ Equation \ref{eq:SS3_97} to calculate the residuals, unless specifically noted otherwise. The absolute residual ellipticity of a single measurement with respect to the sample estimate is then the norm $\left| r_i \right|$.

The difference $\Delta e=\left<\hat e \right> - e$, commonly referred to as simply the bias of the estimator, is formally called the mean-bias $\mu_{\Delta e}$. An estimator is then called asymptotically mean-unbiased, if for an increasing number of estimations $\hat e$, the mean estimate $\mu_{\hat e}$ converges toward the parameter value of the underlying population. This is commonly simply referred to as unbiased. Here we have changed notation from $\left<\hat e \right>$ to $\mu_{\hat e}$, to emphasize the method of determining the central value of a set of estimates.

We do this, because there are other possible definitions of unbiasedness, such as median-unbiasedness, in which case the median estimate $M(\hat e)$ converges toward the true parameter value. By the central limit theorem, it is often appropriate to assume an asymptotically normal distribution of the estimator $\hat e$ (not to be confused with the distribution $P(e)$ of the population parameter $e$), when the number of estimations increases. This validates the general use of mean-unbiasedness. In practice, sample sizes needed for convergence toward a normal estimator distribution can be very large and one should take care when assuming asymptotic normality when making statistical inferences from a few measurements.

The efficiency of an estimator can be defined in terms of its variance. For unbiased estimators, this variance is bounded from below by the Cram\' er-Rao lower bound \citep{Rao_1945,Cramer_1946}, which in short means that there is an absolute maximum efficiency that can be obtained. For some distributions, such as the Gaussian distribution, this limit can be calculated analytically\footnote{We will omit a more detailed discussion, since it's applicable only to certain distributions and not (directly) relevant to this discussion.}. In other cases, it is useful to define a relative efficiency \begin{equation}
\label{eq:efficiency}
\eta_{\hat e} = \frac{\sigma^2_0}{\sigma^2_{\hat e}} \,,
\end{equation} where $\sigma^2_0$ is the variance of a comparison estimator, such as the mean. Then, if for example $\eta_{\hat e} > 1$, the estimator has a lower variance than the mean and is therefore more efficient in finding the central value of the population parameter distribution. An estimator that achieves the Cram\' er-Rao lower bound for all possible parameter values is for this reason also known as a minimum variance estimator.

Again, if the assumption of asymptotic normality is not appropriate, another definition of the scale of distribution of the estimator can be used instead of the variance, such as the median absolute deviation (MAD). In such cases, care should be taken with the coverage of that scale, which is simply the percentage of estimates with lower residuals than the scale. In case of a Gaussian distribution, the standard deviation has a coverage of $68.3\%$. The MAD has, by definition, a coverage of $50\%$.

To avoid comparing apples with oranges, we will use chosen percentiles as scale, so the coverage is defined. For instance, we define the $68.3\%$ scale $s_{68.3}$ as the residual value for which $68.3\%$ of the estimates has an equal or lower residual. In case of asymptotic normality, $s^2_{68.3}$ will converge to the same value as the estimator variance.

We note that we can do this, since in our simulations the true population parameter value $e$ is known\footnote{More accurately, the underlying shear $g$ is known.}. In general, the coverage of a definition of scale is not known, confusing the interpretation of any relative efficiency.

In conclusion, we define the efficiency of an estimator $\hat e$, relative to the mean $\mu_e$, at a certain percentile coverage $p$, as \begin{equation}
\label{eq:rel_eff}
\eta_{\hat e;p} = \frac{s^2_{\mathrm{\mu_e};p}}{s^2_{\hat e;p}} .
\end{equation}

Finally, we label an estimator $\hat e$ as robust (in a qualitative manner), when $\hat e$ retains low or zero bias and high efficiency in a wide range of possible distributions. A robust estimator is desirable, since it makes the choice of estimator for a parameter with unknown distribution more objective. As an example, the mean is optimally efficient in case of a Gaussian parameter distribution, but since the mean has low resistance against departures from Gaussianity (such as outliers), it is not the most robust.

Since we work with relative efficiencies, a conclusive statement about robustness is not straightforward. We will therefore use robustness to indicate that an estimator is equally or more efficient than the Gaussian estimator in most or all cases.

\subsection{Estimators}

We have explored various alternatives for well known estimators, which are optimal under Gaussian assumptions, like the mean and variance. By definition, the mean $\hat e_\mu$, or $\mu_e$, minimizes the variance of the residuals, which makes it a least squares estimator.

In general, optimization estimators are solutions $\hat e$ that minimize a loss function\begin{equation}
\label{loss_f}
S_{\hat e} = \sum_{i} \rho\left(e_i;\hat e \right) \,,
\end{equation}such as $\rho=r_i^2=\left(e_i-\hat e \right)^2$ for the mean.

For this paper, we considered two other optimization estimators, the least absolute deviations estimator (LAD) and the biweight (BI) estimator, and an ordering estimator, namely convex hull peeling (CHP). In section \ref{sec:FMF}, we describe Fourier mode fitting (FM), using a LAD regression approach.


\subsubsection{Least absolute deviations} \label{subsec:LAD}

LAD is an optimization approach where the loss function to be minimized is the sum of the absolute deviations, instead of the commonly used least squares minimization:

\begin{equation}
\label{eq:LAD}
S_\mathrm{LAD}=\sum_i |r_i| .
\end{equation}

In the one dimensional case, this is the median. In more than one dimension, we talk about the marginal median, when in each dimension the median is taken independently, or the spatial median, when minimizing the sum of the distances of measurements to a point. In many practical cases\footnote{Formally speaking: when the norm is strictly convex.}, the spatial median is unique, contrary to the marginal median, which can have multiple solutions. This is one of the reasons we used the spatial median throughout the rest of the paper.

\begin{figure}[h]
\centering
\resizebox{\hsize}{!}{
\includegraphics[angle=90]{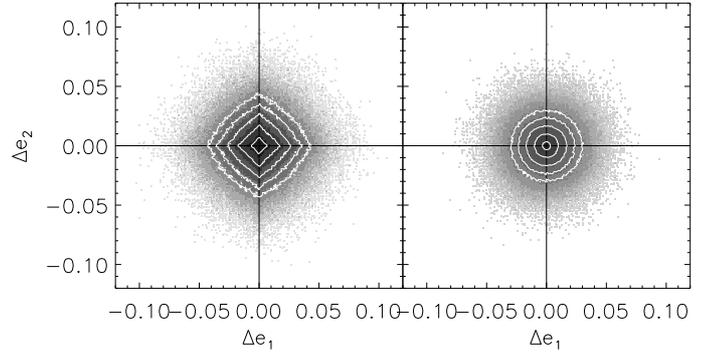}
}
\caption{Comparison of the marginal median (left) to the spatial median, or LAD estimation (right). Plotted are the estimation biases $\Delta e$ for $10^6$ simulation runs, shown as a density in grayscale. Over-plotted are arbitrary contours of increasing density (equal in both plots), to highlight the anisotropy in $P(\Delta e)$. Since $e_1$ and $e_2$ depend on the choice of reference frame, the marginal median introduces an artificial anisotropy. For the LAD estimations, the residual distances $|r_i|$ do not depend on the choice of reference frame.}
\label{MED_example}
\end{figure}

Another reason is that $e_1$ and $e_2$ should not be seen as independent parameters of the shape. An ellipticity is defined by an absolute elongation $|e|$ and a position angle $\theta $. The latter is defined within the context of a chosen frame of reference and therefore so are $e_1$ and $e_2$. In other words, using the marginal median would introduce an artificial anisotropy, as can be seen in Figure \ref{MED_example}.

In concreto, for a set of $(e_{i,1},e_{i,2})$ measurements, the mean $\hat e_\mu$ as an estimator for the net reduced shear $g_1 +ig_2$ minimizes the squared residuals\begin{equation}
\label{eq:S_LSQ}
S_\mu=\sum_i (e_{i,1}-g_1)^2+(e_{i,2}-g_2)^2 .
\end{equation}A LAD estimate minimizes the absolute residuals,\begin{equation}
\label{eq:S_LAD}
S_\mathrm{LAD}=\sum_i \sqrt{(e_{i,1}-g_1)^2+(e_{i,2}-g_2)^2} \,,
\end{equation}which reduces the effect of outliers on the estimate. In one dimension, the LAD estimate arises as the central value maximum likelihood estimator of the Laplace distribution, which has a central cusp and more slowly declining tails.

There is no general analytic solution for LAD optimization. LAD can however be formulated as a linear optimization problem for which several iterative methods exist \citep[e.g., simplex-based methods,][]{Bar_Rob73}. In practical weak shear analyses, convergence is generally rapid.

\subsubsection{The biweight}

An alternative optimization approach is a bi-square weighted loss function \citep{BI_1974}, called the biweight for short, given by \begin{equation}
\label{eq:BIW}
\nabla S_\mathrm{BI}=\sum_i r_i\left( 1-\left(\frac{r_i}{k}\right)^2\right)^2 = 0 \quad \mathrm{for} \quad |r_i|<k \,,
\end{equation} where $r_i=(e_i-\hat e)$ are again the residuals and $k$ is a tuning parameter, usually determined by (an estimate of) the scale of the measured distribution.

A robust choice for $k$ is the median absolute deviation (MAD), setting $k=c\cdot \mathrm{MAD}$, where $c=6.0$ is optimal for estimation of location for a broad range of distributions \citep{BI2_1977}. A common approach is iteratively correcting an initial estimate $M_0$ by the normalized sum in Equation \ref{eq:BIW}: \begin{equation}
\label{eq:BIWform}
M_{n+1}=M_n + \frac{\sum_i r_{i,n}\left( 1-\left(\frac{r_{i,n}}{k}\right)^2\right)^2}{\sum_i \left( 1-\left(\frac{r_{i,n}}{k}\right)^2\right)^2} \,,
\end{equation} which can be interpreted as a normalized weighting of the residuals. In this case, the weight of a certain measurement increases toward the (current) central estimate, which makes this estimator a useful complement to the mean and LAD estimators.

In turn, a robust choice for $M_0$ is the (spatial) median. Note that measurements with residuals $|r_{i,n}|\geq k$ have effectively zero weight, although these points are not `clipped' from the sample, since the residuals can change with each iteration. Convergence usually requires few iterations.

\subsubsection{Convex hull peeling}

\begin{figure}[h]
\centering
\resizebox{\hsize}{!}{
\includegraphics[angle=90]{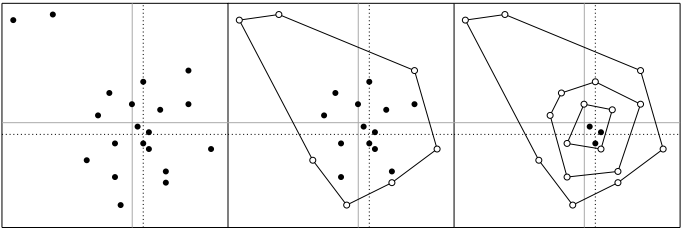}
}
\caption{The method of CHP. The left panel shows a scatter plot with two outliers. The arithmetic mean is shown as a gray, solid line and the dotted line represents the mean without the two outliers. The middle panel shows the convex hull of the set of points, which is then removed from the set. The right panel shows the final result after repeating the process, until the final set of points is equal to its own convex hull.}
\label{CHP_example}
\end{figure}

The convex hull of a set of points $X$ in $\mathbb{R}^n$ can be defined as the intersection of all convex sets in $\mathbb{R}^n$ that contain $X$. Informally put, the convex hull is the smallest subset of points that `surrounds' the rest of the set (see Figure \ref{CHP_example}).

There exist various algorithms for determining the convex hull \citep[e.g.,][]{CHP_alg}. For this paper, we used Delaunay triangulation based on the divide-and-conquer method \citep{CHP_tri}.

In the process of CHP, the convex hull of a set of data points is determined and subsequently `peeled' from the set, after which the process is repeated (Figure \ref{CHP_example}). When the remaining set of points is equal to its own convex hull, the final estimate is determined from these points, for example using the mean or LAD. This makes CHP an ordering approach, much like obtaining the familiar median for the one-dimensional case by sorting the data, instead of optimization\footnote{Indeed, in one dimension, both approaches to the median are the same.}. Among other aspects, it shares the resistance of the median against outliers.

In this paper, we are interested in the use of CHP in the two-dimensional case of $(e_1,e_2)$ measurements in the complex plane, but CHP can be used in higher dimensions as well \citep[see e.g.,][for applications to SDSS quasar data]{CHP_Lee}.

\subsubsection{Weighting and collinearity} \label{CHP_weight}

When using real data, a weighting of ellipticity measurements is necessary to avoid or mitigate effects, such as noise or intrinsic size and ellipticity, that would confuse or bias the estimation of the underlying shear. For LAD and biweight optimization, weighting schemes are readily introduced, analogous to the weighted mean. For CHP, we suggest a possible weighting scheme, analogous to the one-dimensional weighted median, as follows.

The convex hull comprises a set of points in the $(e_{i,1},e_{i,2})$-plane, with $w_i$ the associated weights, given by the measurement pipeline. The minimum weight on the convex hull is then subtracted from these weights, after which all points with updated weight $w_i=0$ are peeled from the sample. Note that this removes at least one point per iteration, but can lead to point-by-point peeling and large computation times. A solution with lesser precision but increased speed would be given by binning the weights in discrete steps.

We also note a possible collinearity problem of multiple ellipticity measurements with finite precision coinciding. In that case, triangulation has no solution. By combining these points into one measurement by combining the weights, this problem is resolved.

\subsection{Fourier mode fitting} \label{sec:FMF}

One can model a signal, in our case a varying ellipticity, over a one-dimensional range $-L<x<L$, writing that signal as a linear superposition of waves, or (Fourier) modes, $A_n \cdot \cos \left( k_nx \pm \phi_n \right)$, where $A_n$ and $\phi_n$ are the amplitude and phase of the signal mode respectively, and $k_n \equiv \frac{n\pi}{L}$ are the wave numbers of the modes, showing the periodicity over the range $2L$. 

It is useful to rewrite this model linearly in its coefficients $a_n \cdot \cos \left( k_nx \right) + b_n \cdot \sin \left( k_nx \right)$, where amplitude and the phase are now given by $A_n^2 = a_n^2 + b_n^2$ and via $\frac{b_n}{a_n}= \tan \left( \phi_n \right)$.

This one-dimensional model is readily extended to two dimensions, by considering that each coefficient depends similarly on $y$, giving $\alpha_{mn;\pm}=\cos \left( k_mx \pm l_ny \right)$ and $\beta_{mn;\pm}=\cos \left( k_mx \pm l_ny \right)$, or \begin{equation}
\label{eq:modes}
\begin{array}{rl}
e(x,y) = \sum_{m,n} & a_{mn} \cos \left( k_mx \right) \cos \left( l_ny \right) \\
                 + & b_{mn} \cos \left( k_mx \right) \sin \left( l_ny \right) \\
                 + & c_{mn} \sin \left( k_mx \right) \cos \left( l_ny \right) \\
                 + & d_{mn} \sin \left( k_mx \right) \sin \left( l_ny \right) \,,
\end{array}
\end{equation} where the wave numbers $k_m$ and $l_n$ represent the spatial frequencies in the $x$ and $y$ directions, respectively. In two dimensions, we make a terminological distinction between a full Fourier mode, as given by Equation \ref{eq:modes}, and the individual waves comprising it. The amplitude of the fluctuations in ellipticity are now given for each mode in $m,n$ by $a_{mn}^2 + b_{mn}^2 + c_{mn}^2 + d_{mn}^2$.

This linear model is fitted in a relatively straightforward manner to a sample of measured or simulated ellipticities. In the absence of noise and for a well-behaved field of view, each wave component of a Fourier mode is independent and can be fitted separately. We will discuss the effect of noise in Section \ref{res:FMF}.

\subsubsection{Applying statistics} \label{sec:apply}

To apply these statistics to a shear field consisting of discrete Fourier modes, which by construction is centered around $e=0$, the ellipticity measurements should be properly weighted by the model of the Fourier mode under consideration. We considered the information carried by an ellipticity measurement, which is proportional to the value of the fitted model $M$, where $M(x,y)$ can for instance be a single wave like $\cos \left( k_mx \right) \cos \left( l_ny \right)$, or a full mode.

Measurements close to the nodes of a wave carry the least information, whereas measurement close to extrema, or antinodes, carry the most amplitude information. We considered that each ellipticity measurement $e_i$ theoretically infers an estimate of the amplitude $A$, where $A\in\{a_{mn},b_{mn},c_{mn},d_{mn} \}$ by ${\hat{A}}_i=e_i\cdot M^{-1}$. In the case of Gaussian variations around the model, that is, measurement error distribution, the information scales as the inverse variance of that distribution, and therefore as the square of the model: \begin{equation}
\label{eq:info}
\hat{A}=\frac{\sum M^2\cdot \frac{e}{M}}{\sum M^2}=\frac{\sum M\cdot e}{\sum M^2} \,,
\end{equation} where we recover the well known analytic LSQ form. This can be seen as an inverse variance weighting based on the model-to-noise ratio. For different error distributions, one can allow a general scaling of the information with the model by $M^n$, and therefore \begin{equation}
\label{eq:geninfo}
\hat{A}=\frac{\sum M^{n-1}\cdot e}{\sum M^n} .
\end{equation} 

\begin{figure*}[h]
\centering
\resizebox{\hsize}{!}{
\includegraphics[angle=90]{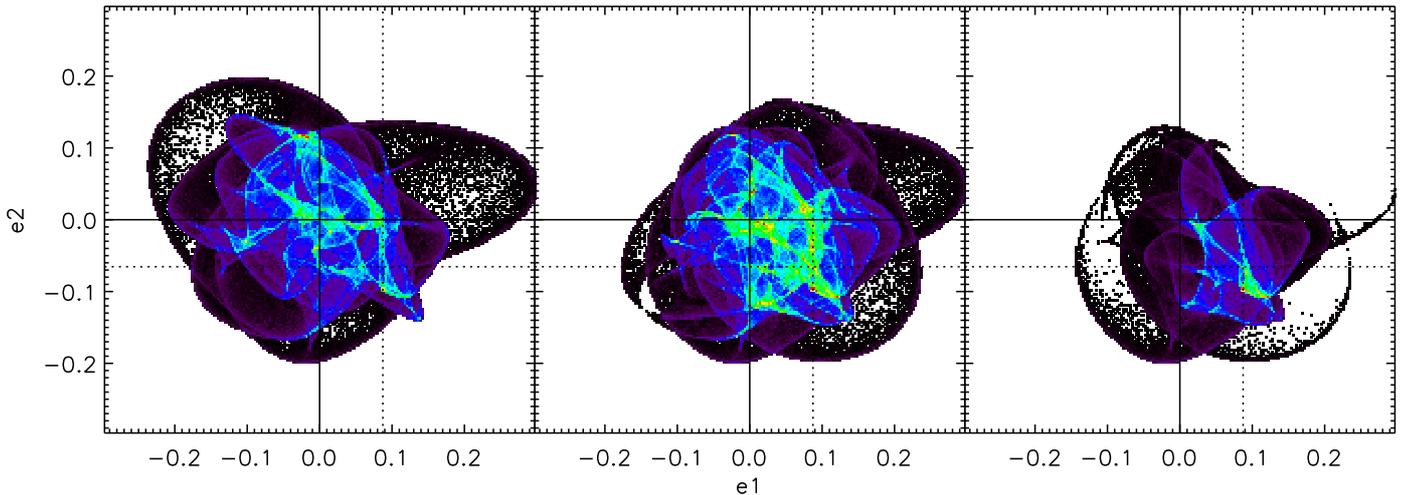}
}
\caption{Ellipticity distribution of a superposition of 9 Fourier modes over the complex $(e_1,e_2)$ plane, where we show how we recover a single $(e_1,e_2)$ amplitude, indicated by the dotted lines. Left: the Fourier modes, centered around $(0,0)$ (normalized number counts). Middle: the $(e_1,e_2)$ points shifted by $\sgn(M)$, with $M$ the model of the amplitude (normalized number counts). Right: the resulting distribution of weights over $(e_1,e_2)$, showing a shift toward the amplitude under consideration.}
\label{fig:shift}
\end{figure*}

For application with our proposed weighting scheme for CHP, it is instructive to view the multiplication by weights as shifting the data points, so the central data point(s) or CHP value matches the amplitude to be estimated. For this purpose, it is practical to write Equation \ref{eq:geninfo} as \begin{equation}
\label{eq:chpinfo}
\hat{A}=\frac{\sum \lvert M\rvert^{n-1}\cdot\sgn(M) \cdot e}{\sum \lvert M\rvert^{n-1}} \cdot\frac{\sum \lvert M\rvert^{n-1}}{\sum M^n}\,,
\end{equation}where the $(e_1,e_2)$ data points are first shifted by $\sgn(M)$, and then weighed by $\lvert M\rvert$, before the weighed estimate is normalized as usual. We show this in Figure \ref{fig:shift}, where we plot the $(e_1,e_2)$ values, the same $(e_1,e_2)$ points shifted by $\sgn(M)$, with in this case $M=\cos \left( k_mx \right) \cos \left( l_ny \right)$, and then the associated distribution of the weights $\lvert M\rvert$ over the complex ellipticity plane as normalized 2D histograms.

\section{Simulations and data}

For this paper, we tested various forms of $P(e)$, using samples of random ellipticities, assumed to be centered around zero, which we sheared by Eq. \ref{eq:SS3_97}. We have used several approaches to obtaining these samples.

Firstly, we simulated a uniform $q$ distribution, which seems to fit real data adequately \citep[e.g.,][]{LML92,Rodriguez_intsh}, without assuming any physical mechanism that would explain this distribution.

Secondly, we modeled background galaxies as randomly orientated triaxial ellipsoids, and derived the projected ellipticities following \citet{Stark77}, using axis ratio distributions fitted to observed ellipticity distributions \citep{LML92}.

In both cases, we compared our results to samples with added Gaussian noise, using real data shape measurement error distributions to simulate the effect of noise.

Thirdly, we sampled real data, using shape measurement catalogs from weak lensing observations.

Finally, we compared these various ellipticity distributions and the results from each estimator to results in case when $P(e)$ follows a Gaussian distribution. We examined the behavior of bias and efficiency of each estimator under the effect of noise, the input shear and the sample size.

\subsection{Simulated ellipticity distributions}

\subsubsection{Uniform samples}

We produced random samples with a uniform $q$-distribution, as an ideal version of the observed distribution of spiral galaxies in for example \citet{LML92,Rodriguez_intsh}, henceforth referred to as a uniform sample. We used an axis ratio cut-off of $q\approx 0.2$ to account for a finite galaxy thickness, following \citet{LML92}, which gives rise to standard deviations in each ellipticity component of $\sigma_{e}\approx 0.25$, comparable to the samples drawn from data.

\begin{figure}[h]
\centering
\resizebox{\hsize}{!}{
\includegraphics{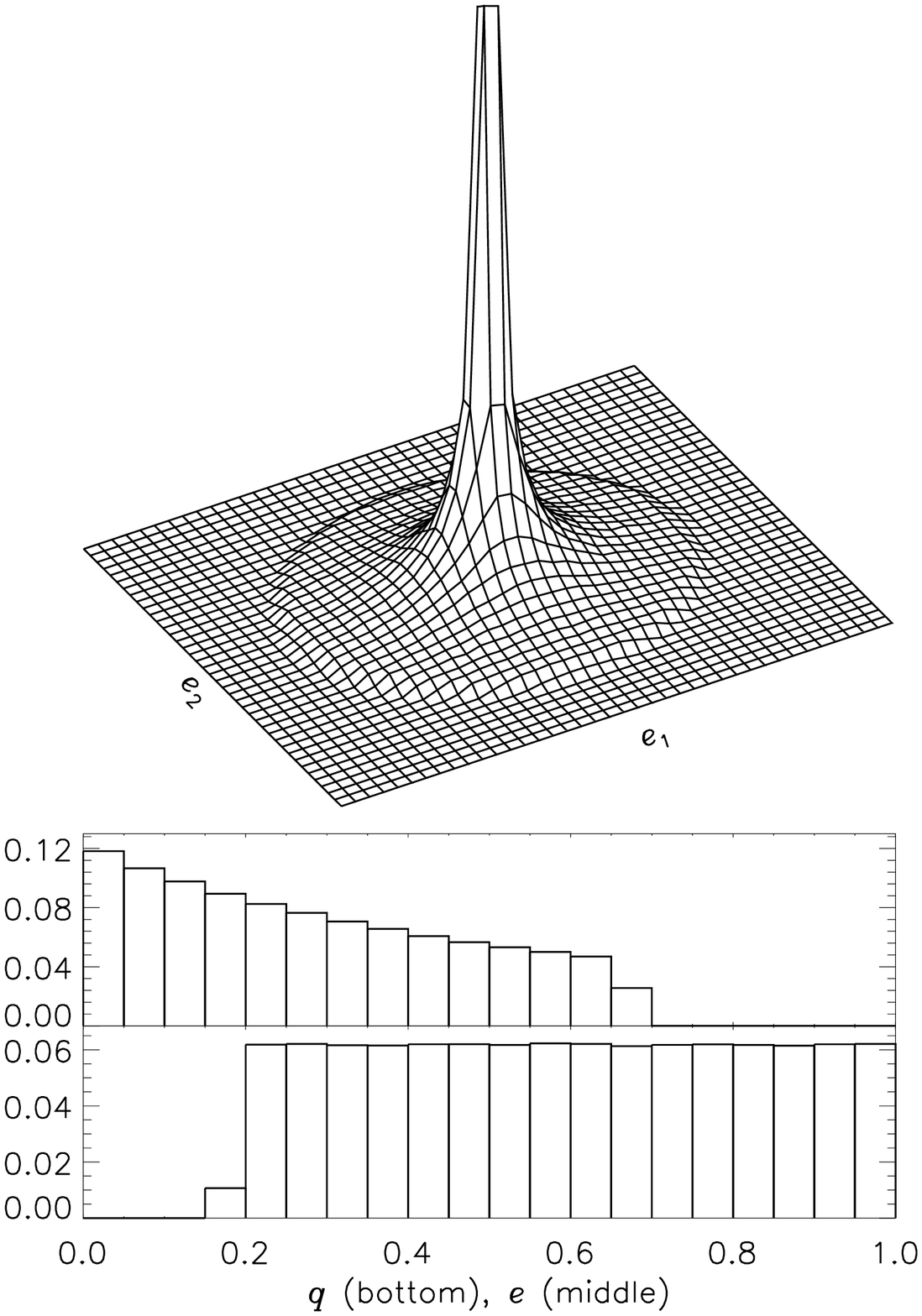}
}
\caption{Ellipticity distributions for a uniform axis ratio distribution. Top: a 2D histogram of ellipticities. Middle: histogram of the absolute ellipticity $|e|$. Bottom: histogram of the ellipse axis ratio $q$. A cut-off near $q\approx 0.2$ is suggested by observations and produces standard deviations in each ellipticity component of $\sigma_{e}\approx 0.25$, comparable to most survey shape measurement catalogs.}
\label{fig:dist_flat}
\end{figure}

The resulting axis ratio and ellipticity distributions are shown in Figure \ref{fig:dist_flat}

\subsubsection{Projected ellipsoids}

A triaxial ellipsoid with axes $\tilde{a}\ge\tilde{b}\ge\tilde{c}\ge 0$ can be described by \begin{equation}
\label{eq:ellipsoid}
\left( cx\right)^2+\left( \frac{cy}{b}\right)^2+\left( z\right)^2=\mathrm{constant} \,,
\end{equation} with $b=\tilde{b} /\tilde{a}$ and $c=\tilde{c} /\tilde{a}$. As given by \citet{Stark77}, such an ellipsoid is seen as an ellipse in projection, given by \begin{equation}
\label{eq:projection}
(j/f)x'^2+2(k/f)x'y'+(l/f)y'^2=\mathrm{constant} \,,
\end{equation} where $(x',y')$ are the coordinates in the projection plane and \begin{subequations}
\label{eq:subeq}
\begin{align}
f & \equiv c^2 \sin^2 \theta \sin^2 \varphi + (c/b)^2 \sin^2 \theta \cos^2 \varphi + \cos^2 \theta \,, \\
j & \equiv c^2 (c/b)^2 \sin^2 \theta + c^2 \cos^2 \varphi \cos^2 \theta + (c/b)^2 \sin^2 \varphi \cos^2 \theta \,, \\
k & \equiv ((c/b)^2-c^2) \sin \varphi \cos \varphi \cos \theta \,, \\
l & \equiv c^2\sin^2 \varphi + (c/b)^2 \cos^2 \varphi \,,
\end{align}
\end{subequations}with $\varphi$ and $\theta$ the first two orientation angles of the ellipsoid.

For simulations of projected ellipsoids, we assumed Gaussian distributions for $b$ and $c$, following \citet{LML92}. For elliptical galaxies, we used $b=0.95$ and $c=0.55$ with standard deviations $\sigma_{b} =0.35$ and $\sigma_{c} =0.2$. For disk galaxies, we used $b=1.00$ and $c=0.25$ with standard deviations $\sigma_{b} =0.13$ and $\sigma_{c} =0.12$.

\begin{figure*}[h]
\centering
\resizebox{\hsize}{!}{
\includegraphics{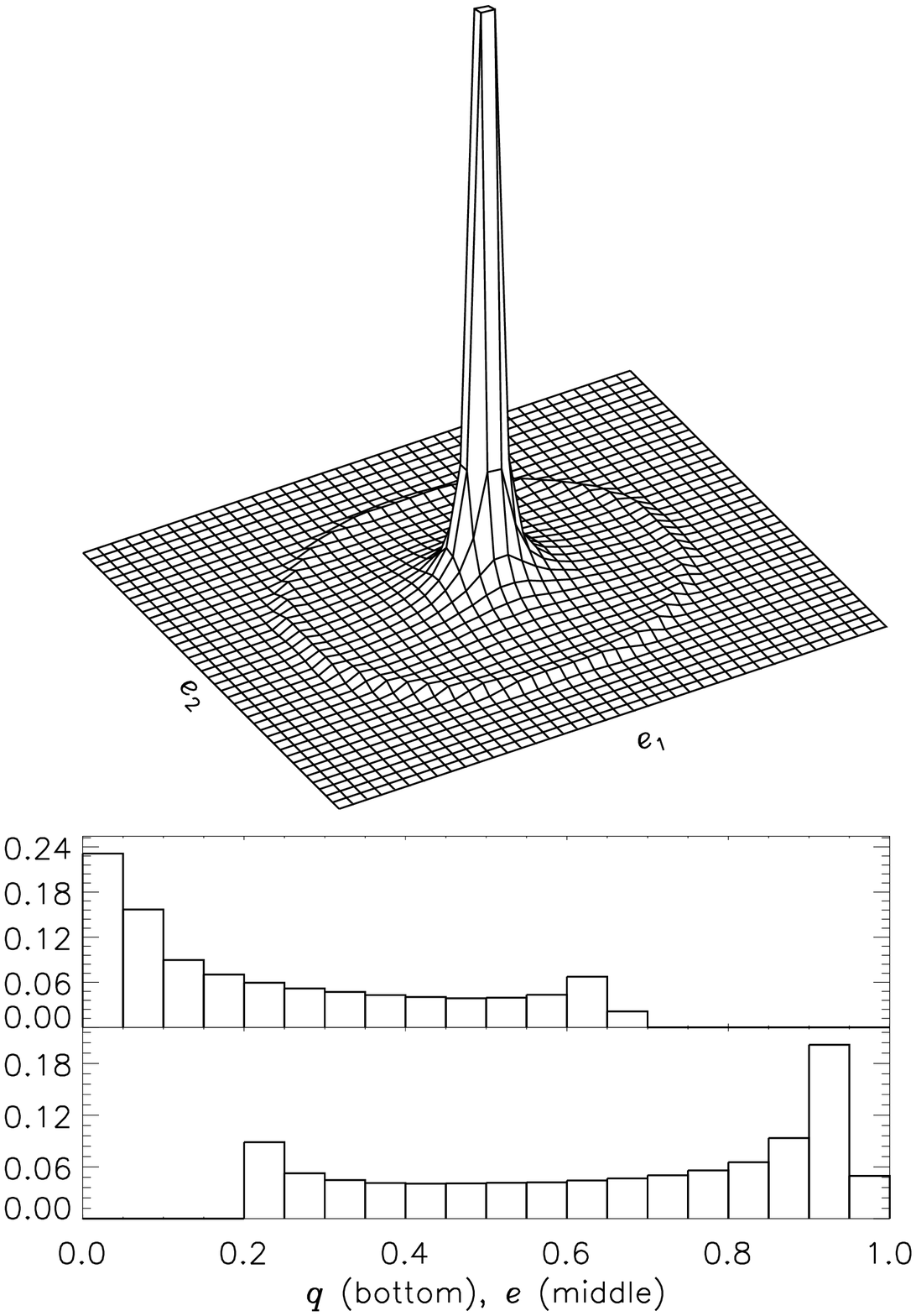}
\includegraphics{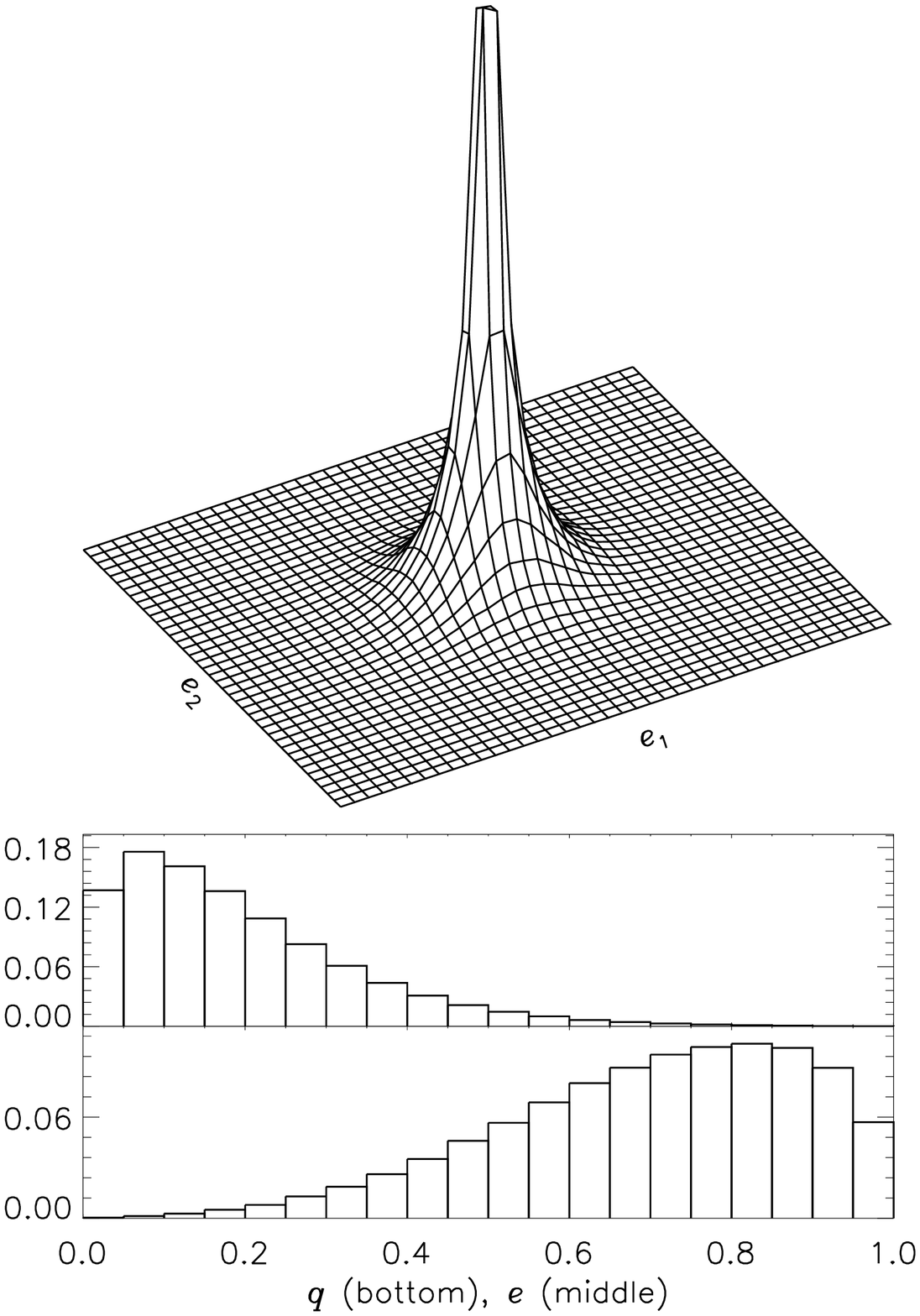}
}
\caption{Ellipticity and axis ratio distributions for distribution of projected ellipsoids. Left: disk galaxies. Right: elliptical galaxies. Top: a 2D histogram of ellipticities. Note that the ring-like feature in the left panel is the result of a finite disk thickness. Middle: histogram of the absolute ellipticity $|e|$. Bottom: histogram of the ellipse axis ratio $q$.}
\label{fig:dist_stark}
\end{figure*}

The axis ratio was then recovered via \begin{equation}
\label{eq:revproj}
q = \sqrt{\frac{j+l-\sqrt{\left( j-l \right)^2+4k^2}}{j+l+\sqrt{\left( j-l \right)^2+4k^2}}} \,,
\end{equation} and the ellipticity through Equation \ref{eq:axis}. The orientation angles of the ellipsoids were randomly distributed. The resulting axis ratio and ellipticity distributions are shown in Figure \ref{fig:dist_stark}.

We will refer to these simulated samples as disk and elliptical samples. We also used combined samples with a disk to elliptical ratio derived from the CFHTLenS catalog. (See Table \ref{tab:CFHTLenS}.)

\subsection{Data: CFHTLenS}

We used data from Canada-France-Hawaii Lensing Survey \citep[][CFHTLenS]{CFHTLenS_Heymans12}. The CFHTLenS survey analysis combined weak lensing data processing with {\sc theli} \citep{THELI_Erben05, CARS_Erben09, CFHTLenS_Erben13}, shear measurement with {\em lens}fit \citep{Lensfit_Miller07, CFHTLenS_Miller13, Lensfit_Kitching08}, and Bayesian photometric redshift measurement \citep[BPZ,][]{BPZ_Benitez00,BPZ_Coe06} with PSF-matched photometry \citep{CFHTLenS_Hildebrandt12}. A full systematic error analysis of the shear measurements in combination with the photometric redshifts is presented in \citet{CFHTLenS_Heymans12}, with additional error analyses of the photometric redshift measurements presented in \citet{CFHTLenS_Benjamin13}.

For our analyses, we selected 4.2 million objects that are well determined and resolved ({\em lens}fit $\mathrm{fitclass}=0$, non-zero {\em lens}fit weight, $\mathrm{star\_flag}=0$, ${\tt CLASS\_STAR}\le 0.5$). We excluded objects that lie within a mask, with the exception of large, conservative masks around relatively faint stars and stellar haloes \citep[$\mathrm{MASK}\le 1$, see][]{CFHTLenS_Erben13}.

The CFHTLenS shape catalog is not an exact representation of the ellipticity distribution of the observed galaxy population, as it includes measurement noise present in any real data set. Selecting sources on {\em lens}fit weight $w$ or signal-to-noise ratio $\nu_{\mathrm{SN}}$ could on the other hand introduce selection biases in the galaxy population we wanted to to study. We decided to use two sets of sources: the complete set, described above, to optimally sample the complete source population, and a conservative subset with $w \ge 15$ and $\nu_{\mathrm{SN}} \ge 20$, to reduce the uncertainty in observed ellipticity, at the possible cost of a bias in the selection.

For both sets, we split these sources by BPZ spectral type into red ($T_\mathrm{BPZ}<1.5$) and blue ($1.5<T_\mathrm{BPZ}<3.95$) galaxies, with a further division between Sbc ($1.5<T_\mathrm{BPZ}<2.5$) and Scd ($2.5<T_\mathrm{BPZ}<3.95$). We found that our conservative selection reduced the number of galaxies to roughly 25\%, almost independent of spectral bin for $T_\mathrm{BPZ}<3.1$. For higher spectral types, the subset decreased linearly to roughly 10\% for the highest spectral bin, which was an indication that our selection did indeed introduce a modest sample bias.

\begin{figure*}[h]
\centering
\resizebox{\hsize}{!}{
\includegraphics{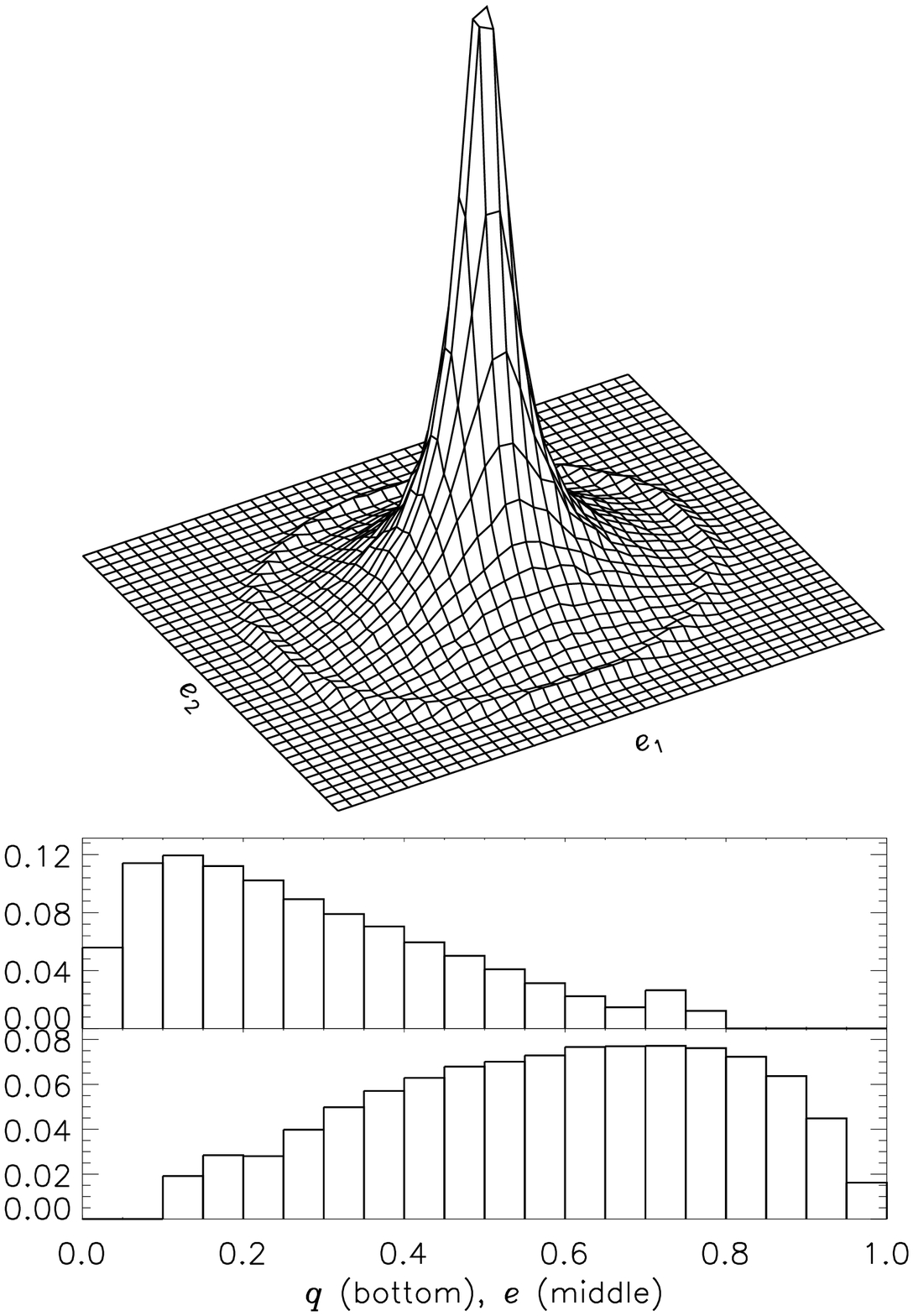}
\includegraphics{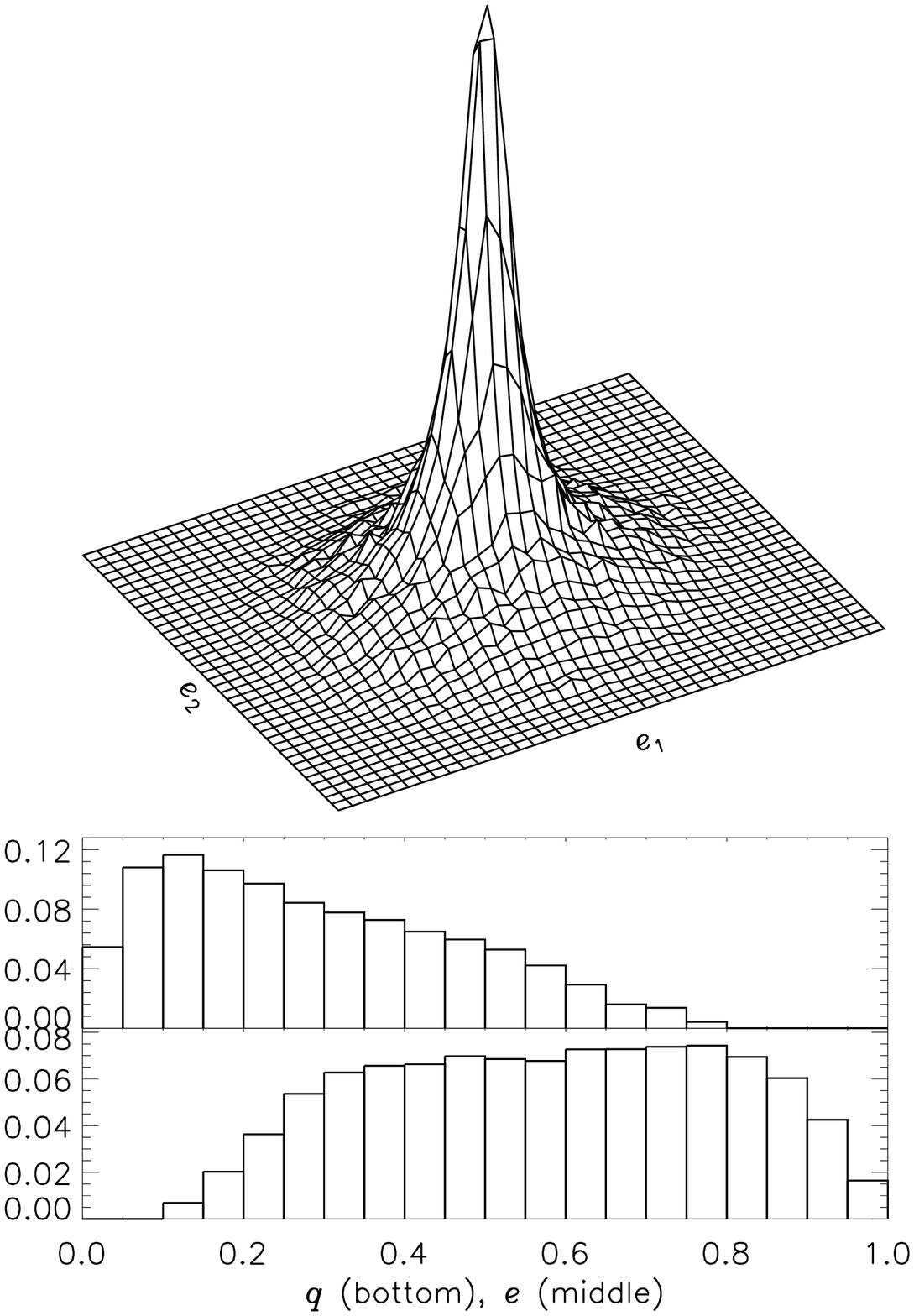}
}
\caption{Ellipticity and axis ratio distributions for CFHTLenS blue (left), red (right). Top: a 2D histogram of ellipticities. Note that the ring-like feature at $e\approx 0.8$ is due to noisy outliers forced to a maximum $e$ by the shape measurement pipeline, but see also Figure \ref{fig:dist_stark}. Middle: histogram of the absolute ellipticity $|e|$. Bottom: histogram of the ellipse axis ratio $q$.}
\label{fig:dist_cfhtls_col}
\end{figure*}

Table \ref{tab:CFHTLenS} gives an overview of the selected CFHTLenS data, while Figure \ref{fig:dist_cfhtls_col} shows the respective distributions.

\begin{table}[h]
\caption{Overview of the CFHTLenS data used. Column 1 gives the division between BPZ spectral type (red: $T_\mathrm{BPZ}<1.5$, blue: $1.5<T_\mathrm{BPZ}<3.95$, Sbc: $1.5<T_\mathrm{BPZ}<2.5$, Scd: $2.5<T_\mathrm{BPZ}<3.95$). Column 2 gives the number $N$ of objects selected. Column 3 gives the 1D Gaussian ellipticity standard deviation $\sigma_e$, using both ellipticity components after bias correction. In parentheses, we give $N$ and $\sigma_e$ for sources with $w \ge 15$ and $\nu_{\mathrm{SN}} \ge 20$.}
\label{tab:CFHTLenS}
\centering
\begin{tabular}{lcccc}
\hline
\hline
Color & $N$ & & $\sigma_e$ & \\
\hline
All    & 4216334  & (912828) & 0.286 & (0.242) \\
Red    & 553633   & (151939) & 0.267 & (0.242) \\
Blue   & 3662701  & (760889) & 0.289 & (0.242) \\
Sbc    & 870295   & (219929) & 0.294 & (0.262) \\
Scd    & 2792406  & (540960) & 0.288 & (0.232) \\ 
\hline                                                           
\end{tabular}                                                    
\end{table}

We drew random subsets from the selected CFHTLenS ellipticities, which we then sheared by Eq. \ref{eq:SS3_97}. This introduced the implicit assumption that, after the bias corrections described in \citet{CFHTLenS_Heymans12} and \citet{CFHTLenS_Miller13}, the central ellipticity was zero, and these random subsets were approximately drawn from an unsheared, noise-free background galaxy population.

\subsection{Simulated noise} \label{data:noise}

In any realistic shape measurement catalog, ellipticities not only have shape noise due to a finite intrinsic distribution, but suffer from measurement uncertainties as well. For this reason, we wanted to study the effect of noise or our simulated, noiseless ellipticity samples.

Measurement uncertainties depend primarily on pixel noise and therefore vary with image size and brightness. This means that errors on the ellipticities are not drawn from a single distribution. To mimic the effect of a skewed composite error distribution for our simulated samples, we randomly sampled the CFHTLenS weight $w$.

\citet{CFHTLenS_Miller13} calculated an approximately inverse-variance weight using the width of the ellipticity likelihood surface by\begin{equation}
\label{eq:cfhtls_w}
w=\left[\frac{\sigma^2_ee^2_\mathrm{max}}{e^2_\mathrm{max}+2\sigma^2_e}+\sigma^2_\mathrm{pop}\right]^{-1} \,,
\end{equation}where $\sigma^2_e$ is the variance in ellipticity of the likelihood surface, $\sigma^2_\mathrm{pop}$ is the ellipticity variance of the galaxy population, and $e_\mathrm{max}$ is a maximum ellipticity, to reflect a finite edge-on disk thickness.

Using $e_\mathrm{max}=0.804$ from \citet{CFHTLenS_Miller13} and refining $\sigma^2_\mathrm{pop}\approx 0.242$ using the CFHTLenS catalog itself\footnote{\citet{CFHTLenS_Miller13} cite $\sigma^2_\mathrm{pop}=0.255$ as prior, but this would lead to a negative $\sigma^2_e$ for the maximum weight in the CFHTLenS catalog.}, we obtained a distribution in ellipticity variance $\sigma^2_e$ for each $w$. From this, we produced noise by assuming a Gaussian distribution with the ellipticity as mean and $\sigma^2_e$ as variance.

\subsubsection{Estimation of errors}

To assess errors on bias and efficiency from our simulations, we simply divided our simulations randomly in smaller subsets and determine the statistical variations, assuming $t$-distributions. While this approach may seem to lack finesse compared to a full bootstrap, the significance of our results is high enough for a proof of concept.

\section{Results} \label{results}                                             
                                                                 
\subsection{Central value estimation} \label{res:central}

For each sample type, we produced $10^4$ random samples of $100$ ellipticities, which we distorted by an absolute reduced shear of $g=0.2$, and determined relative efficiencies and possible biases. We then assessed the effect of varying the shear and the sample size.

\subsubsection{Asymmetry and bias} \label{res:bias}

Ideally, an estimator should be unbiased in the absence of noise. For the mean, this is the case \citep{SS3_97}, but since the effect of shear on intrinsic ellipticities in non-linear, the resulting, observed ellipticity distribution $P(\tilde{e})$ is asymmetric, or skewed, which can lead to mean-biases for various estimators.

\begin{figure}[h]
\centering
\resizebox{\hsize}{!}{
\includegraphics[angle=90]{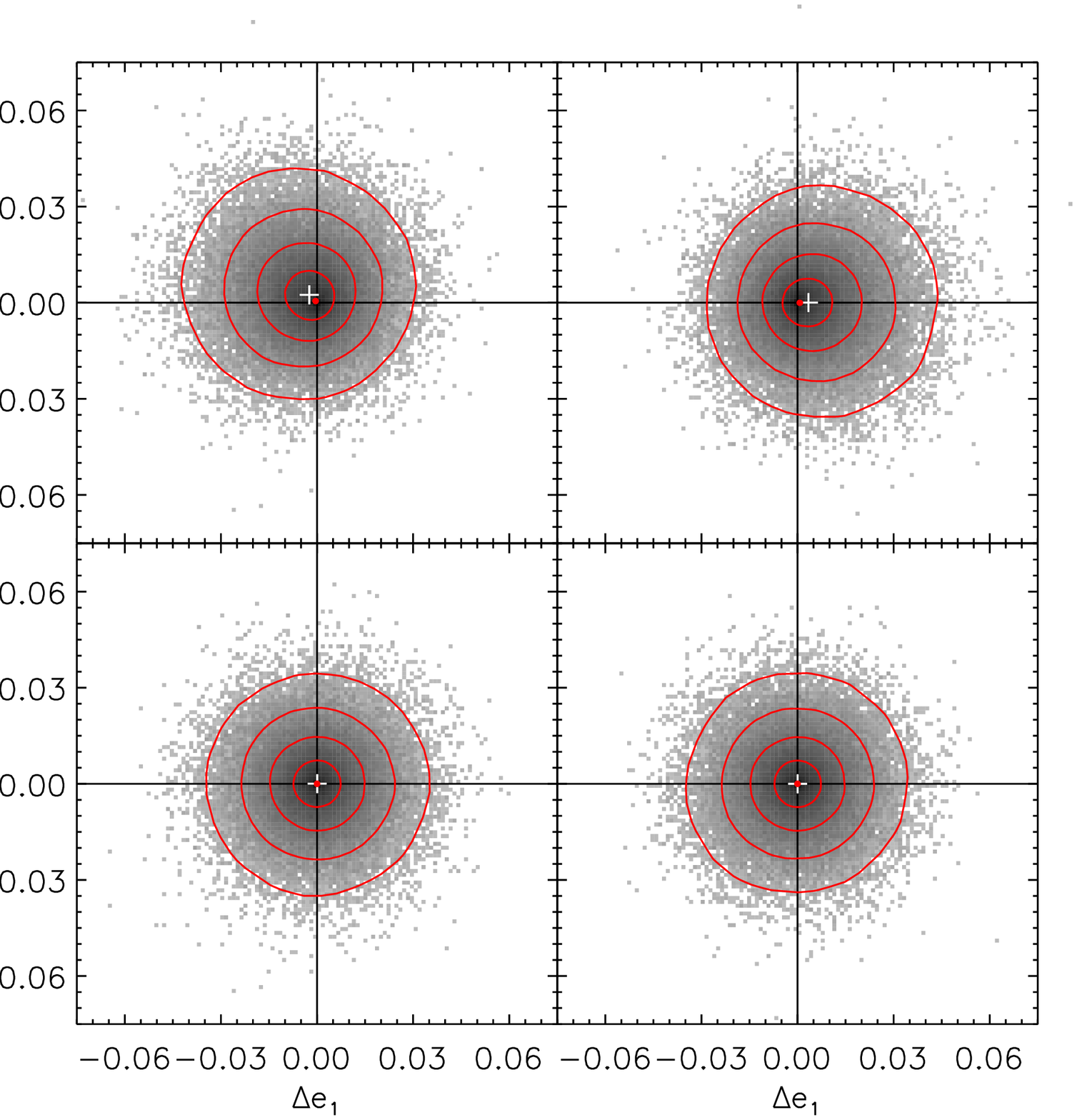}
}
\caption{The skewed $\hat e_{\mathrm {CHP}}$ distribution as an example of the effect of asymmetry in a sheared ellipticity distribution. Plotted are the estimation biases $\Delta e$ for $10^5$ simulation runs, shown as a density in grayscale. Over-plotted are the convex hulls at approximately $38.3\%$, $68.3\%$, $86.6\%$, and $95.4\%$ coverage. The mean of the distribution is shown as a white plus. Top: estimation biases $\Delta e_{\mathrm{ CHP}}$ for samples with an underlying shear of $g=-0.21+0.21i$ (left) and $g=0.3$ (right). Note that these estimator distributions are effectively mean-biased, because they are skewed, but still centered around $\Delta e=0$, as indicated by the CHP estimation of the distributions. Bottom: $\Delta e_{\mathrm {CHP}}$ for the same samples, after iteratively de-shearing the samples until the final CHP estimate vanishes. These iterations remove asymptotic mean-bias and increase efficiency.}
\label{peel_bias}
\end{figure}

In Figure \ref{peel_bias}, we show this effect on the CHP estimator for $g=0.3$ in two directions. The distribution of the CHP estimator is clearly skewed, as shown by the convex hulls plotted, when the coverage within the current hull is equal to approximately\footnote{CHP is a discrete and not a continuous process, but this effect is negligible for $10^4$ estimates.} $38.3\%$, $68.3\%$, $86.6\%$, and $95.4\%$. We note that this leads to a mean-biasedness, according to definition, but the center of the estimator distribution $P(\hat{e}_{\mathrm {CHP}})$ seems significantly less biased. In other words, the CHP estimator seems `CHP-unbiased'.

A solution to this skewness in the estimator distribution, in the absence of noise, is iteratively improving estimates by correcting the observed ellipticities $P(\tilde{e})$ by the estimated shear, using Equation \ref{eq:SS3_97}, and then determining the updated residuals. We call this process of iteratively correcting the sample by the current estimate `de-shearing' (or `de-$g$'). Figure \ref{peel_bias} shows how this symmetrized the estimator distribution $P(\hat{e})$, and slightly improved the efficiency as well (see section \ref{res:efficiencies}). The latter seemed to be the case even for the mean $\hat{e}_\mu$ as estimator, but the difference was not statistically significant.

In presence of noise, the mean is a biased estimator \citep{Noise_Melchior12}. Given that in reality systematic noise is always present, a form of bias is unavoidable, since the noise distribution is different\footnote{Intrinsically, the effect of noise is symmetric, but the effect on a sample of sheared ellipticities depends on the shape measurement pipeline, as noted in \citet{Noise_Melchior12}.} from the (skewed) ellipticity distribution (See Figure \ref{fig:mapping}). This means that our method of de-shearing would introduce a noise bias for precisely the same reason, since we would not properly correct the asymmetry in the distribution.

We compared the results for simulated projected ellipsoids with and without simulated noise in Figure \ref{fig:est_mean_bias} to assess the effect. In the appendix, we quantified the observed multiplicative bias in the form \begin{equation}
\label{eq:bias}
e_{\mathrm{fit}}=(1+m)e_{\mathrm{in}} \,,
\end{equation} where $e$ stands for $e_{1,2}$, and summarize the results in Table \ref{tab:bias}.

Without de-shearing, only the mean is a mean-unbiased estimator. We noted that all estimation methods could be made mean-unbiased in the noise-free case, when including de-shearing, but showed a mean-bias in the presence of noise, as expected. For the biweight estimator $\hat{e}_{\mathrm {BI}}$, this was (within statistical significance) the same bias as for the mean. For the LAD and CHP estimators $\hat{e}_{\mathrm {LAD}}$ and $\hat{e}_{\mathrm {CHP}}$, the biases were significantly reduced, up to $\sim 30\%$, to below percent level for realistic weak shear.

\begin{figure*}[h]
\centering
\resizebox{\hsize}{!}{
\includegraphics[angle=90]{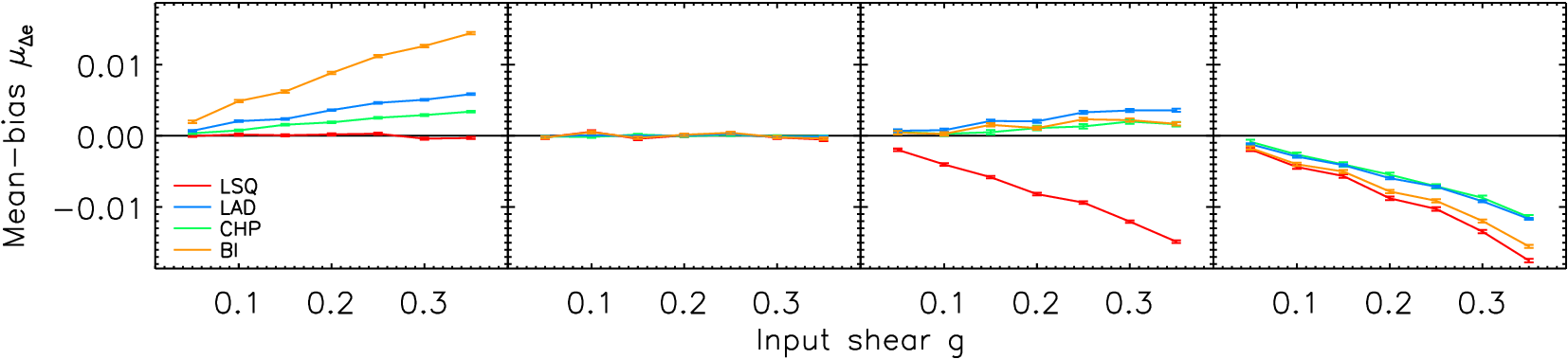}
}
\caption{Estimator mean-bias as a function of input shear for realistic combinations of simulated disk and elliptical samples, using projected ellipsoids. From left to right: all estimators without noise, without noise after iteratively de-shearing the samples, all estimators with noise, and with noise and after iteratively de-shearing the samples. Color coding: $\hat{e}_{\mu}$ (red), $\hat{e}_{\mathrm {LAD}}$ (blue), $\hat{e}_{\mathrm {CHP}}$ (green) and $\hat{e}_{\mathrm {BI}}$ (yellow).}
\label{fig:est_mean_bias}
\end{figure*}

This decrease in bias can be explained by realizing that the observed, sheared ellipticity distribution is skewed, but the location of the central peak of intrinsically round background sources is still an unbiased estimator of the underlying shear (which can be deduced from Equation \ref{eq:SS3_97} and Figure \ref{fig:mapping}). It is the bias in determining the location of this peak that introduces the bias in the shear estimate. Likewise, the effect of noise changes the observed ellipticity distribution, but does not affect the location of that peak. Estimators that are more sensitive to a central cusp or peak in the distribution and less to high ellipticities in the tail, such as $\hat{e}_{\mathrm {LAD}}$ and $\hat{e}_{\mathrm {CHP}}$, will therefore introduce a lower mean-bias.

We compared these results to the mean-bias in the upper panels of Figure \ref{peel_bias} and the observation that the central peak of the estimator distribution is in fact located at $\Delta e\approx 0$. We found that the mean-bias arose due to the asymmetry in the estimator distribution and the CHP-bias vanished, unaffected by noise.

\subsubsection{Estimator efficiencies} \label{res:efficiencies}

In Tables \ref{tab:LAD}, \ref{tab:BI}, and \ref{tab:CHP} in the appendix, we summarized the full results for the relative efficiencies of each estimator. We applied de-shearing and note that this improves the efficiencies marginally at a similar marginal cost to the bias. We determined relative efficiencies for coverages of $25\%$, $50\%$, and $75\%$, corresponding to the MAD and the first and third quartiles, and $38.3\%$, $68.3\%$, $86.6\%$, and $95.4\%$, which would correspond to steps of $0.5\sigma$ in case of a Gaussian distribution with variance $\sigma^2$.

\begin{figure*}[h]
\centering
\resizebox{\hsize}{!}{
\includegraphics[angle=90]{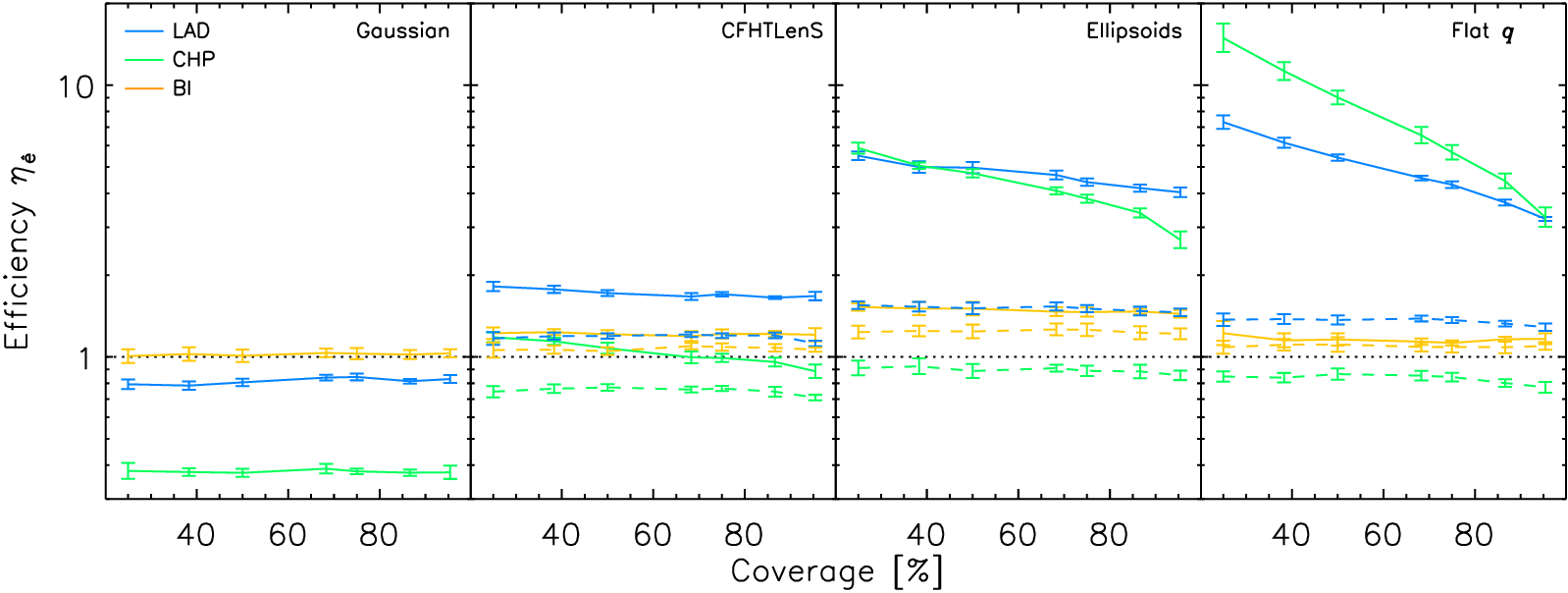}
}
\caption{Relative efficiencies each estimator plotted at different coverages. From left to right: Relative efficiencies in case of a Gaussian $P(e)$ distribution, the CFHTLenS catalog $P(e)$ distribution, a combination of disk and elliptical distributions using projected ellipsoids, and a uniform $q$ distribution. Color coding: relative efficiencies for $\hat e_{\mathrm{LAD}}$ (blue), $\hat e_{\mathrm{BI}}$ (yellow), and $\hat e_{\mathrm{CHP}}$ (green). Solid lines: simulated samples without noise or using the CFHTLenS conservative subset. Dashed lines: including noise or using the complete CFHTLenS set.}
\label{fig:eff_cover}
\end{figure*}

In Figure \ref{fig:eff_cover}, we plot these results for a few distributions, namely Gaussian, uniform $q$, a combination of disk and elliptical projections and the conservative CFHTLenS catalog samples. We also plot the results for the samples with added noise and the full CFHTLenS samples.

Not all estimators reached asymptotic normality. Especially CHP converged slower toward normality in the tails of the distribution, that is, at higher coverage. For LAD, this is noticeable mostly for the uniform $q$ distribution.

The biweight is the most robust, as its relative efficiency doesn't vary much across distributions. The biweight relative efficiency is however quite low, which means that this estimator offers little improvement. Even when $P(e)$ follows a Gaussian distribution, $\eta_{\mathrm BI}$ is not significantly better or worse than the traditional mean.

\begin{figure*}[h]
\centering
\resizebox{\hsize}{!}{
\includegraphics[angle=90]{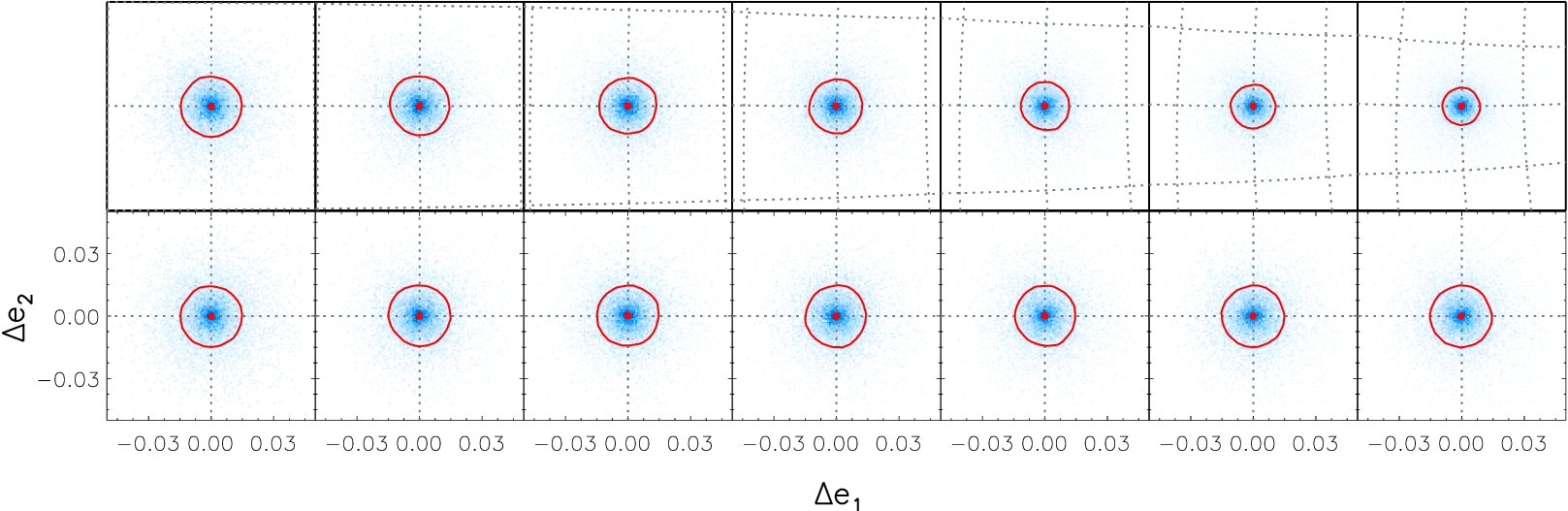}
}
\caption{Efficiency for an arbitrary estimator and sample type versus input shear, ranging from $g=0.05$ (left) to $g=0.35$. Upper: simple difference $\hat{e}-g$ between estimates and input shear, with $s_{68.3}$ over-plotted. Lower: $\Delta e$, as defined by Equation \ref{eq:def_diff}, with $s_{68.3}$ over-plotted.}
\label{spread_shear}
\end{figure*}

Our results show that estimator efficiency is independent of input shear. This is the case, when we define the individual estimate biases similarly to the residuals, as noted in Section \ref{bias_eff_rob}, that is, not as the difference $\hat{e}-g$, but as the extra shear needed over the input shear $g$ to reach this difference, as determined by Equation \ref{eq:SS3_97}:\begin{equation}
\label{eq:def_diff}
\Delta e = \frac{\hat{e}-g}{1-g^*\hat{e}} \,,
\end{equation}with $g^*$ the complex conjugate of the input shear $g$ of the simulations. Using that definition, this independence is demonstrated Figure \ref{spread_shear}.

As an aside: since the mean of the CHP estimator is displaced from the center, this necessarily increases the distribution scale. A more proper way to compare the scale with symmetric distributions would be comparing the surface within the convex hull at a certain coverage, as $s^2$ is a measure of the (circular) surface around $\hat{e}$ inside that scale. In this sense, efficiency is a figure of merit. We have not done so in this paper, which means the $\eta_{\mathrm{CHP}}$ are slightly underestimated, but not significantly.

In Figure \ref{fig:eff_sample}, we show the results for different samples sizes. In Table \ref{tab:eff_sample} in the appendix, we summarize the quantitative results. In the limit of very small sample sizes, the difference between the various estimators is expected to vanish. We note that a potential improvement over the mean estimator remains even for a sample size of $N=10$.

\begin{figure*}[h]
\centering
\resizebox{\hsize}{!}{
\includegraphics[angle=90]{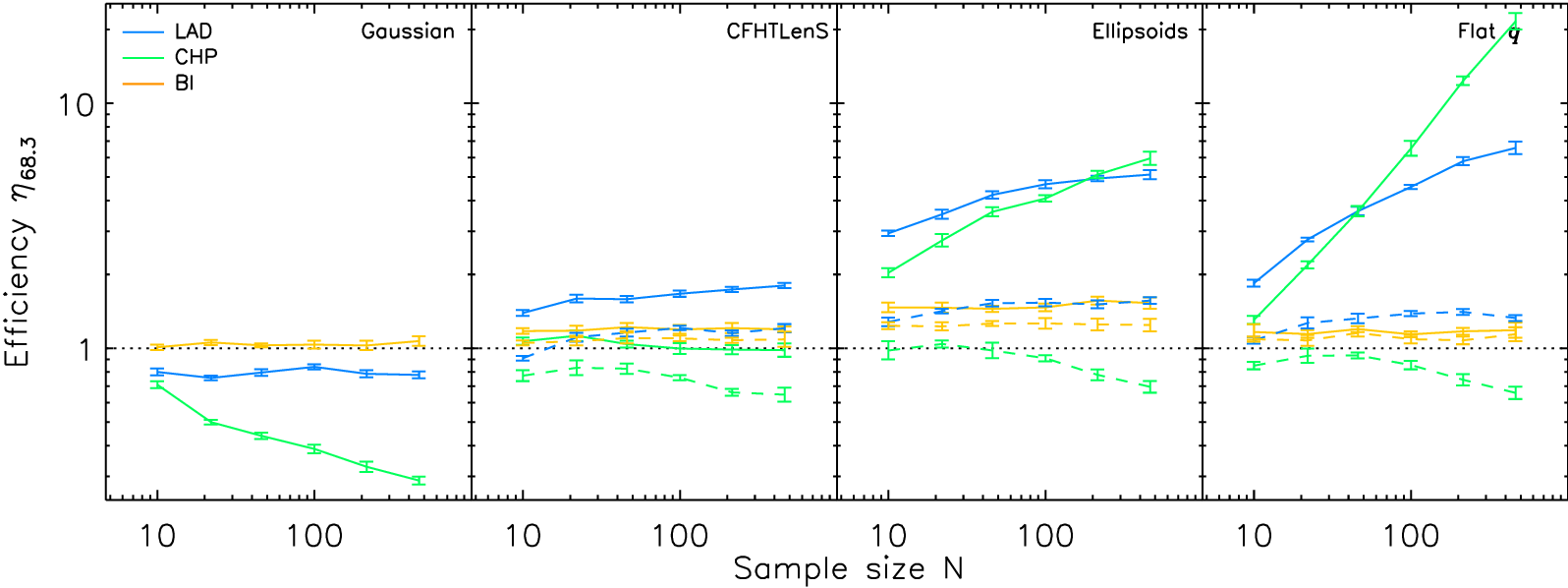}
}
\caption{Relative efficiencies $\eta_{68.3}$ plotted against sample sizes. From left to right: Relative efficiencies in case of a Gaussian $P(e)$ distribution, the CFHTLenS catalog $P(e)$ distribution, a combination of disk and elliptical distributions using projected ellipsoids, and a uniform $q$ distribution. Color coding: relative efficiencies for $\hat e_{\mathrm{LAD}}$ (blue), $\hat e_{\mathrm{BI}}$ (yellow), and $\hat e_{\mathrm{CHP}}$ (green). Solid lines: simulated samples without noise or using the CFHTLenS conservative subset. Dashed lines: including noise or using the complete CFHTLenS set.}
\label{fig:eff_sample}
\end{figure*}

\subsection{Fourier mode fitting} \label{res:FMF}

For samples of a combination of disk and elliptical distributions using projected ellipsoids, we produced $10^3$ random square fields with $10^3$ simulated ellipticities. For comparison, the average number of selected sources in a CFHTLenS field is roughly $2.5\cdot 10^4$, ranging from 9525 to 37767, or $5.3\cdot 10^3$, ranging from 2111 to 9525 for the more conservative sample.

Using Equation \ref{eq:SS3_97}, we distorted these intrinsic ellipticities by the total shear pattern of one or more full modes (as defined in Equation \ref{eq:modes}), then applied simulated measurement noise (as described in Section \ref{data:noise}) as a final step.

We fitted amplitudes per individual wave using LSQ, LAD and CHP, and per mode using LSQ and LAD by simultaneously fitting all four amplitudes. We then determined relative efficiencies and possible biases of the recovered amplitudes in the same way as in Section \ref{res:central}.

\begin{figure*}[h]
\centering
\resizebox{\hsize}{!}{
\includegraphics[angle=90]{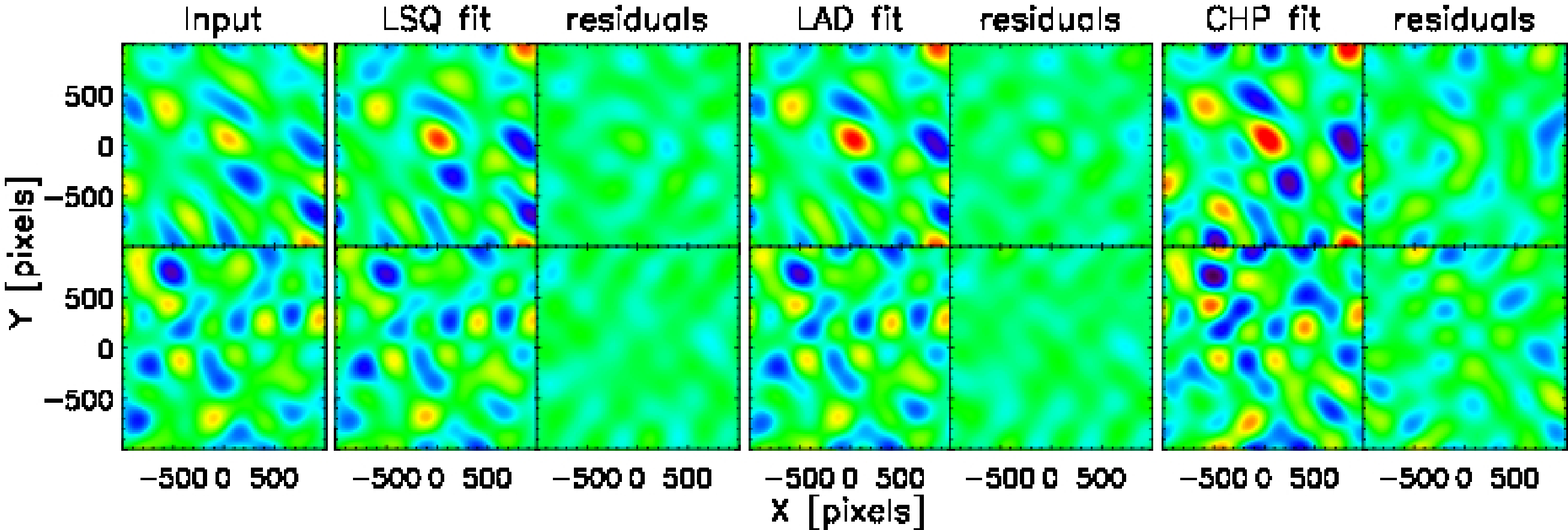}
}
\caption{Shear field residuals when fitting 16 different modes, using simulated projected ellipsoids as intrinsic shapes, and including additional Gaussian noise. From left to right: input shear, LSQ fit and residuals, LAD fit and residuals, CHP it and residuals. Upper and lower row show $e_1$ and $e_2$ respectively. The color scale is the same in all plots for comparison and ranges between $-0.247\le e_{1,2}\le 0.247$. Residuals for this realization vary between $\pm 0.075$,$\pm 0.066$ and $\pm 0.14$, respectively. {\em Note: image quality reduced for faster arxiv download. Original image available on request or in the journal version.}}
\label{fig:shear_field}
\end{figure*}

In Figure \ref{fig:shear_field}, we show the fitted shear field for a single realization, using in this case $10^4$ simulated ellipticities. We fitted 16 different modes individually, using LSQ and LAD, and 64 individual amplitudes using LSQ, LAD, and CHP, and found the amplitude residuals, $\left(\mathcal{O}\left(10^{-3} \right)\right)$, to be two orders of magnitude less than the input values, which were constrained to $g\leq 0.25$ for peak values at positive interference. Residuals in $|e|$ for this realization varied between $\pm 0.075$ for LSQ,$\pm 0.066$ for LAD and $\pm 0.14$ for CHP.

\subsubsection{Bias and efficiency} \label{res:FMF_bi_eff}

\begin{figure*}[h]
\centering
\resizebox{\hsize}{!}{
\includegraphics[angle=90]{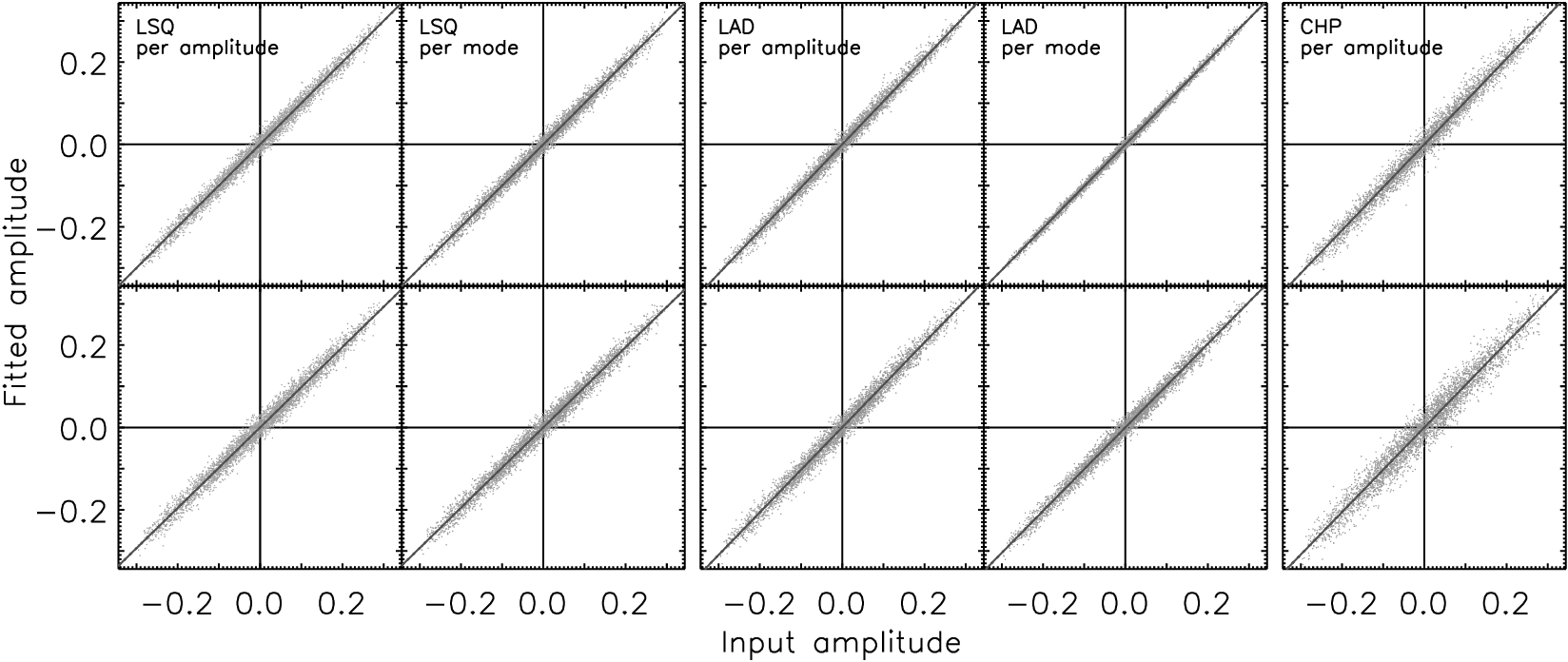}
}
\caption{Consistency of the estimated Fourier amplitudes as defined in Equation \ref{eq:modes}. Plotted are the input amplitudes $a_{mn}$, $b_{mn}$, $c_{mn}$, $d_{mn}$ versus their estimates (Section \ref{sec:apply}) for LSQ (left), LAD (middle) AND CHP (right). Top row: only simulated intrinsic shapes based on projected ellipsoids. Bottom row: the same, with added noise. Over plotted are the best-fitting mean-bias, as defined in Equation \ref{eq:bias_FMF}.}
\label{fig:amp_abs}
\end{figure*}

The results from Sections \ref{res:bias} and \ref{res:efficiencies} carry over to estimates of Fourier amplitudes for LSQ and LAD. We found fitted values with standard deviations of the order of $10^{-3}$ for individual amplitudes. In Figure \ref{fig:amp_abs}, we show the consistency of the fitted values. 

Over plotted in Figure \ref{fig:amp_abs} are the best-fitting mean-bias, defined similar to Equation \ref{eq:bias} as\begin{equation}
\label{eq:bias_FMF}
a_{\mathrm{est}}=(1+m)a_{\mathrm{in}}\,,
\end{equation} where $a$ stands for $a_{mn}$, $b_{mn}$, $c_{mn}$, and $d_{mn}$ as defined in Equation \ref{eq:modes}. The uncertainties are too small to be visible. In Table \ref{tab:bias_FMF} in the appendix, we give the quantitative results.

Similar to the results shown in Figure \ref{fig:est_mean_bias}, LSQ underestimates Fourier amplitudes by a few percent in the presence of noise. For LAD, we found an improvement on bias by $\sim 20\%$ in the presence of noise, when iteratively de-shearing the sample. Likewise, fitting for LAD without de-shearing slightly overestimated the amplitudes, again comparable to Figure \ref{fig:est_mean_bias}.

We note that in this case, adding noise did not seem to have a significant effect when fitting per mode. In most cases, we did notice a significant increase in bias when fitting per single amplitude. We did not see a change in bias between LSQ per mode and per amplitude.

We also found a slightly higher relative efficiency of $\eta_{68.3}=1.09\pm 0.07$ for LSQ and $\eta_{68.3}=1.47\pm 0.09$ for LAD, when fitting per mode, with or without added noise. It is not surprising that a model with four parameters (amplitudes) fits the estimates better than a model with one parameter, but the difference of this effect between LSQ and LAD is noteworthy.

Since CHP doesn't fit a model to the data, but rather orders the $(e_1,e_2)$ data points, there is no straightforward way to fit four amplitudes simultaneously with the necessary weighting (Section \ref{CHP_weight}). We have not explored this option further in this paper.

The CHP estimator performs consistently, that is, convergent around the input values, but with a significant lower efficiency than for central value estimation of a cloud of $(e_1,e_2)$ data points (Section \ref{res:central}). This is to be expected, since CHP is particularly sensitive to a (central) cusp in the distribution of data points. By shearing the intrinsic ellipticities by a model that varies over the field of view, as shown in Figure \ref{fig:shift}, this peak will be smeared out, decreasing the effectiveness of CHP.

In conclusion, CHP is consistently the most sensitive to the central cuspiness of a distribution. The results of this section do serve as a proof of concept for applying alternative statistics to an observed field of weak shear measurements.

\section{Conclusions and summary} \label{Summary}

\subsection{Optimal estimators}

Our main conclusion is that to evaluate a statistical estimator, one must be willing to look beyond the canonical terms of mean-bias and the Gaussian variance as efficiency. We have shown that these commonly used meta-analysis instruments do not always properly reflect how well weak shear estimator values are constrained around the true underlying shear values.

By discussing the statistical definitions and observing the behavior of estimators for various ellipticity distributions, we have proposed ways of comparing various estimators motivated by statistical theory. The conclusions of that comparison are as follows:

Since the central peak of the intrinsic ellipticity distribution $P(e)$ is an unbiased tracer of the underlying shear, we find that the LAD and CHP estimators are less biased and more efficient than the standard mean.

When iteratively de-shearing the ellipticity sample by the estimated shear, the LAD estimator can reach a sub-percent bias for typical weak shear values, including noise. LAD is generally the most efficient of all estimators considered, potentially reducing uncertainties by more than $50\%$ for samples simulated using a model of projected triaxial ellipsoids.

The CHP estimator is in terms of its mean-bias less affected by noise, as compared to the mean and, to a lesser extend, LAD. In fact, since the estimator distribution $P(\hat{e})$ is not symmetric, the actual center of that distribution, as opposed to the mean of that distribution, is unbiased in the presence of noise, within statistical significance. This makes CHP an important consideration, but it is less straightforward for adaptation for regression and requires careful assessment of uncertainties. Furthermore, CHP is computationally more demanding. In the presence of Gaussian noise, CHP is slightly less efficient than the mean (Figure \ref{fig:eff_cover}, panels 3 and 4), but defining efficiency in terms of a figure of merit can reduce this drawback compared to the gain in bias, as proposed in Section \ref{res:efficiencies}.

\subsection{Direct Fourier mode fitting}

Applying different statistics to fitting individual Fourier modes to the shear field directly, we found results consistent with our previous conclusions.

We have shown that the Fourier amplitudes can be recovered with sub-percent accuracy and a minimal bias, which is an important proof of concept. Since the periodic variations in underlying shear effectively smooth the central peak of the intrinsic ellipticity distribution, the gain in efficiency is slightly less for LAD and significantly less for CHP. It is possible that an alternative to our weighting scheme for FMF with CHP could improve results. At this point, the method of CHP seems more applicable to samples of expected (roughly) constant shear, for example when measuring tangential shear around a gravitational lens candidate in bins of distance.

We have also shown that the shear field can be recovered reliably, with residuals an the order of magnitude less than the variations of the shear over the field of view for LSQ and LAD, using $10^3$ sources, which is conservative compared to a typical single CFHTLenS field.

\subsection{Future considerations and possible applications}

We have discussed alternative statistics for inference of shear from samples of background sources with various intrinsic ellipticity distributions, proposing methods that could improve biases and uncertainties arising from the shape noise. It is important to consider our results within the broader context of other sources of systematics, as mentioned in our introduction.

Firstly, our results for shape noise assume trustworthy shape measurements, not only per source, but also considering the effect of systematics in the shape measurement pipelines on the reproduced ellipticity distribution as a whole: the recovery of a central peak, the distribution of outliers, among others. Examples are the effect of constraining ellipticities to a certain `physical' maximum \citep[e.g., $e_\mathrm{max}=0.804$ for {\em lensfit},][]{CFHTLenS_Miller13}, as we see in Figure \ref{dist_cfhtls}, or conversely, the unphysical outliers with $|e|> 1.0$ arising from dividing two noisy quantities (often when correcting for the point spread function, or PSF), affecting the tails of the distribution. Any features in the recovered shape distribution could affect bias and efficiency of the statistic used. Optimizing statistics will place more stringent demands on shape measurements than performing excellent `on average'. Even methods that avoid individual shape measurements \citep{Bernstein_2014}, an ensemble inferred reduced shear could improve by considering the intrinsic shape distribution. Secondly, even with an accurately measured shape distribution, there will remain sources of systematic error in other steps of a cosmological analysis, as noted in our introduction. These effects still form a necessary part in a weak lensing analysis, but leave our statistical conclusions unaffected.

As survey sizes and image qualities increase, so will the demands on constraining systematic effects to a sub-dominant level, as described in for example \citet{KIDS_DR2} and \citet{HSC_DR1} for the KiDS DR2 and HSC DR1, respectively. At the same time, it will be interesting to see measured ellipticity distributions converge as more sources are observed with higher signal-to-noise and measured with higher fidelity, due to increased depth of imaging, image quality and PSF control.

For now, we have given a proof of concept for alternative statistics in two cases: a sample of ellipticities with one underlying shear and the recovery of individual Fourier modes of the shear variation over a field of view. The first part has important applications when inferring a shear profile around lenses, both in recovering an accurate, less biased estimate and smaller error bars or confidence intervals. For the second part: since the amplitudes are well constrained by fitting individual Fourier modes, this provides a possible method toward estimation of the power spectrum. Furthermore, the shear field can be recovered in terms of its Fourier amplitudes, providing a powerful analytic model for mass reconstruction, without the need for smoothed gridding and incorporating variations in background source densities and estimated measurement uncertainties.

\begin{acknowledgements}

We thank the anonymous referee for useful comments and suggestions, which improved readability and reproducibility of this paper and its results.

MS acknowledges support from the Netherlands Organization for Scientific Research (NWO).

This work is based on observations obtained with MegaPrime/MegaCam, a joint project of CFHT and CEA/DAPNIA, at the Canada-France-Hawaii Telescope (CFHT) which is operated by the National Research Council (NRC) of Canada, the Institut National des Sciences de l'Univers of the Centre National de la Recherche Scientifique (CNRS) of France, and the University of Hawaii. This research used the facilities of the Canadian Astronomy Data Centre operated by the National Research Council of Canada with the support of the Canadian Space Agency. CFHTLenS data processing was made possible thanks to significant computing support from the NSERC Research Tools and Instruments grant program.

\end{acknowledgements}

\bibliographystyle{aa}
\bibliography{ShearStats.bib}

\begin{thebibliography}{65}
\expandafter\ifx\csname natexlab\endcsname\relax\def\natexlab#1{#1}\fi

\bibitem[{Barrodale \& Roberts(1973)}]{Bar_Rob73}
Barrodale, I. \& Roberts, F. D.~K. 1973, SIAM Journal on Numerical Analysis,
  10, 839

\bibitem[{{Bartelmann} \& {Schneider}(2001)}]{Review_Bartelmann_2001}
{Bartelmann}, M. \& {Schneider}, P. 2001, \physrep, 340, 291

\bibitem[{{Beaton} \& {Tukey}(1974)}]{BI_1974}
{Beaton}, A.~E. \& {Tukey}, J.~W. 1974, Outliers in Statistical Data (New York:
  Wiley)

\bibitem[{{Ben{\'{\i}}tez}(2000)}]{BPZ_Benitez00}
{Ben{\'{\i}}tez}, N. 2000, \apj, 536, 571

\bibitem[{{Benjamin} {et~al.}(2013){Benjamin}, {Van Waerbeke}, {Heymans},
  {Kilbinger}, {Erben}, {Hildebrandt}, {Hoekstra}, {Kitching}, {Mellier},
  {Miller}, {Rowe}, {Schrabback}, {Simpson}, {Coupon}, {Fu},
  {Harnois-D{\'e}raps}, {Hudson}, {Kuijken}, {Semboloni}, {Vafaei}, \&
  {Velander}}]{CFHTLenS_Benjamin13}
{Benjamin}, J., {Van Waerbeke}, L., {Heymans}, C., {et~al.} 2013, \mnras, 431,
  1547

\bibitem[{{Bernstein}(2010)}]{Bernstein_2010}
{Bernstein}, G.~M. 2010, \mnras, 406, 2793

\bibitem[{{Bernstein} \& {Armstrong}(2014)}]{Bernstein_2014}
{Bernstein}, G.~M. \& {Armstrong}, R. 2014, \mnras, 438, 1880

\bibitem[{{Bernstein} \& {Jarvis}(2002)}]{BJ02}
{Bernstein}, G.~M. \& {Jarvis}, M. 2002, \aj, 123, 583

\bibitem[{{Bonnet} \& {Mellier}(1995)}]{Bonnet_1995}
{Bonnet}, H. \& {Mellier}, Y. 1995, \aap, 303, 331

\bibitem[{{Bridle} {et~al.}(2010){Bridle}, {Balan}, {Bethge}, {Gentile},
  {Harmeling}, {Heymans}, {Hirsch}, {Hosseini}, {Jarvis}, {Kirk}, {Kitching},
  {Kuijken}, {Lewis}, {Paulin-Henriksson}, {Sch{\"o}lkopf}, {Velander},
  {Voigt}, {Witherick}, {Amara}, {Bernstein}, {Courbin}, {Gill}, {Heavens},
  {Mandelbaum}, {Massey}, {Moghaddam}, {Rassat}, {R{\'e}fr{\'e}gier}, {Rhodes},
  {Schrabback}, {Shawe-Taylor}, {Shmakova}, {van Waerbeke}, \&
  {Wittman}}]{GREAT08_2010}
{Bridle}, S., {Balan}, S.~T., {Bethge}, M., {et~al.} 2010, \mnras, 405, 2044

\bibitem[{{Coe} {et~al.}(2006){Coe}, {Ben{\'{\i}}tez}, {S{\'a}nchez}, {Jee},
  {Bouwens}, \& {Ford}}]{BPZ_Coe06}
{Coe}, D., {Ben{\'{\i}}tez}, N., {S{\'a}nchez}, S.~F., {et~al.} 2006, \aj, 132,
  926

\bibitem[{Cram{\'e}r(1946)}]{Cramer_1946}
Cram{\'e}r, H. 1946, Mathematical Methods of Statistics, Princeton mathematical
  series (Princeton University Press)

\bibitem[{{Dark Energy Survey Collaboration} {et~al.}(2016){Dark Energy Survey
  Collaboration}, {Abbott}, {Abdalla}, {Aleksi{\'c}}, {Allam}, {Amara},
  {Bacon}, {Balbinot}, {Banerji}, {Bechtol}, {Benoit-L{\'e}vy}, {Bernstein},
  {Bertin}, {Blazek}, {Bonnett}, {Bridle}, {Brooks}, {Brunner}, {Buckley-Geer},
  {Burke}, {Caminha}, {Capozzi}, {Carlsen}, {Carnero-Rosell}, {Carollo},
  {Carrasco-Kind}, {Carretero}, {Castander}, {Clerkin}, {Collett}, {Conselice},
  {Crocce}, {Cunha}, {D'Andrea}, {da Costa}, {Davis}, {Desai}, {Diehl},
  {Dietrich}, {Dodelson}, {Doel}, {Drlica-Wagner}, {Estrada}, {Etherington},
  {Evrard}, {Fabbri}, {Finley}, {Flaugher}, {Foley}, {Fosalba}, {Frieman},
  {Garc{\'{\i}}a-Bellido}, {Gaztanaga}, {Gerdes}, {Giannantonio}, {Goldstein},
  {Gruen}, {Gruendl}, {Guarnieri}, {Gutierrez}, {Hartley}, {Honscheid}, {Jain},
  {James}, {Jeltema}, {Jouvel}, {Kessler}, {King}, {Kirk}, {Kron}, {Kuehn},
  {Kuropatkin}, {Lahav}, {Li}, {Lima}, {Lin}, {Maia}, {Makler}, {Manera},
  {Maraston}, {Marshall}, {Martini}, {McMahon}, {Melchior}, {Merson}, {Miller},
  {Miquel}, {Mohr}, {Morice-Atkinson}, {Naidoo}, {Neilsen}, {Nichol}, {Nord},
  {Ogando}, {Ostrovski}, {Palmese}, {Papadopoulos}, {Peiris}, {Peoples},
  {Percival}, {Plazas}, {Reed}, {Refregier}, {Romer}, {Roodman}, {Ross},
  {Rozo}, {Rykoff}, {Sadeh}, {Sako}, {S{\'a}nchez}, {Sanchez}, {Santiago},
  {Scarpine}, {Schubnell}, {Sevilla-Noarbe}, {Sheldon}, {Smith}, {Smith},
  {Soares-Santos}, {Sobreira}, {Soumagnac}, {Suchyta}, {Sullivan}, {Swanson},
  {Tarle}, {Thaler}, {Thomas}, {Thomas}, {Tucker}, {Vieira}, {Vikram},
  {Walker}, {Wechsler}, {Weller}, {Wester}, {Whiteway}, {Wilcox}, {Yanny},
  {Zhang}, \& {Zuntz}}]{DES_2016}
{Dark Energy Survey Collaboration}, {Abbott}, T., {Abdalla}, F.~B., {et~al.}
  2016, \mnras, 460, 1270

\bibitem[{{de Jong} {et~al.}(2013){de Jong}, {Verdoes Kleijn}, {Kuijken}, \&
  {Valentijn}}]{KIDS_2013}
{de Jong}, J.~T.~A., {Verdoes Kleijn}, G.~A., {Kuijken}, K.~H., \& {Valentijn},
  E.~A. 2013, Experimental Astronomy, 35, 25

\bibitem[{{Erben} {et~al.}(2009){Erben}, {Hildebrandt}, {Lerchster}, {Hudelot},
  {Benjamin}, {van Waerbeke}, {Schrabback}, {Brimioulle}, {Cordes}, {Dietrich},
  {Holhjem}, {Schirmer}, \& {Schneider}}]{CARS_Erben09}
{Erben}, T., {Hildebrandt}, H., {Lerchster}, M., {et~al.} 2009, \aap, 493, 1197

\bibitem[{{Erben} {et~al.}(2013){Erben}, {Hildebrandt}, {Miller}, {van
  Waerbeke}, {Heymans}, {Hoekstra}, {Kitching}, {Mellier}, {Benjamin}, {Blake},
  {Bonnett}, {Cordes}, {Coupon}, {Fu}, {Gavazzi}, {Gillis}, {Grocutt}, {Gwyn},
  {Holhjem}, {Hudson}, {Kilbinger}, {Kuijken}, {Milkeraitis}, {Rowe},
  {Schrabback}, {Semboloni}, {Simon}, {Smit}, {Toader}, {Vafaei}, {van Uitert},
  \& {Velander}}]{CFHTLenS_Erben13}
{Erben}, T., {Hildebrandt}, H., {Miller}, L., {et~al.} 2013, \mnras, 433, 2545

\bibitem[{{Erben} {et~al.}(2005){Erben}, {Schirmer}, {Dietrich}, {Cordes},
  {Haberzettl}, {Hetterscheidt}, {Hildebrandt}, {Schmithuesen}, {Schneider},
  {Simon}, {Deul}, {Hook}, {Kaiser}, {Radovich}, {Benoist}, {Nonino}, {Olsen},
  {Prandoni}, {Wichmann}, {Zaggia}, {Bomans}, {Dettmar}, \&
  {Miralles}}]{THELI_Erben05}
{Erben}, T., {Schirmer}, M., {Dietrich}, J.~P., {et~al.} 2005, Astronomische
  Nachrichten, 326, 432

\bibitem[{{Ford} {et~al.}(2012){Ford}, {Hildebrandt}, {Van Waerbeke},
  {Leauthaud}, {Capak}, {Finoguenov}, {Tanaka}, {George}, \&
  {Rhodes}}]{Mag_Ford_2012}
{Ford}, J., {Hildebrandt}, H., {Van Waerbeke}, L., {et~al.} 2012, \apj, 754,
  143

\bibitem[{{Herbonnet} {et~al.}(2017){Herbonnet}, {Buddendiek}, \&
  {Kuijken}}]{Herbonnet_2016}
{Herbonnet}, R., {Buddendiek}, A., \& {Kuijken}, K. 2017, \aap, 599, A73

\bibitem[{{Heymans} {et~al.}(2012{\natexlab{a}}){Heymans}, {Rowe}, {Hoekstra},
  {Miller}, {Erben}, {Kitching}, \& {van Waerbeke}}]{CFHTLenS_Rowe12}
{Heymans}, C., {Rowe}, B., {Hoekstra}, H., {et~al.} 2012{\natexlab{a}}, \mnras,
  421, 381

\bibitem[{{Heymans} {et~al.}(2006){Heymans}, {Van Waerbeke}, {Bacon}, {Berge},
  {Bernstein}, {Bertin}, {Bridle}, {Brown}, {Clowe}, {Dahle}, {Erben}, {Gray},
  {Hetterscheidt}, {Hoekstra}, {Hudelot}, {Jarvis}, {Kuijken}, {Margoniner},
  {Massey}, {Mellier}, {Nakajima}, {Refregier}, {Rhodes}, {Schrabback}, \&
  {Wittman}}]{STEP_2006}
{Heymans}, C., {Van Waerbeke}, L., {Bacon}, D., {et~al.} 2006, \mnras, 368,
  1323

\bibitem[{{Heymans} {et~al.}(2012{\natexlab{b}}){Heymans}, {Van Waerbeke},
  {Miller}, {Erben}, {Hildebrandt}, {Hoekstra}, {Kitching}, {Mellier}, {Simon},
  {Bonnett}, {Coupon}, {Fu}, {Harnois D{\'e}raps}, {Hudson}, {Kilbinger},
  {Kuijken}, {Rowe}, {Schrabback}, {Semboloni}, {van Uitert}, {Vafaei}, \&
  {Velander}}]{CFHTLenS_Heymans12}
{Heymans}, C., {Van Waerbeke}, L., {Miller}, L., {et~al.} 2012{\natexlab{b}},
  \mnras, 427, 146

\bibitem[{{Hildebrandt} {et~al.}(2016){Hildebrandt}, {Choi}, {Heymans},
  {Blake}, {Erben}, {Miller}, {Nakajima}, {van Waerbeke}, {Viola},
  {Buddendiek}, {Harnois-D{\'e}raps}, {Hojjati}, {Joachimi}, {Joudaki},
  {Kitching}, {Wolf}, {Gwyn}, {Johnson}, {Kuijken}, {Sheikhbahaee}, {Tudorica},
  \& {Yee}}]{RCSLenS_2016}
{Hildebrandt}, H., {Choi}, A., {Heymans}, C., {et~al.} 2016, \mnras, 463, 635

\bibitem[{{Hildebrandt} {et~al.}(2012){Hildebrandt}, {Erben}, {Kuijken}, {van
  Waerbeke}, {Heymans}, {Coupon}, {Benjamin}, {Bonnett}, {Fu}, {Hoekstra},
  {Kitching}, {Mellier}, {Miller}, {Velander}, {Hudson}, {Rowe}, {Schrabback},
  {Semboloni}, \& {Ben{\'{\i}}tez}}]{CFHTLenS_Hildebrandt12}
{Hildebrandt}, H., {Erben}, T., {Kuijken}, K., {et~al.} 2012, \mnras, 421, 2355

\bibitem[{{Hildebrandt} {et~al.}(2011){Hildebrandt}, {Muzzin}, {Erben},
  {Hoekstra}, {Kuijken}, {Surace}, {van Waerbeke}, {Wilson}, \&
  {Yee}}]{Mag_Hildebrandt_2011}
{Hildebrandt}, H., {Muzzin}, A., {Erben}, T., {et~al.} 2011, \apjl, 733, L30

\bibitem[{{Hildebrandt} {et~al.}(2009){Hildebrandt}, {van Waerbeke}, \&
  {Erben}}]{Mag_Hildebrandt_2009}
{Hildebrandt}, H., {van Waerbeke}, L., \& {Erben}, T. 2009, \aap, 507, 683

\bibitem[{{Hirata} \& {Seljak}(2003)}]{HS03}
{Hirata}, C. \& {Seljak}, U. 2003, \mnras, 343, 459

\bibitem[{{Hirata} {et~al.}(2004){Hirata}, {Mandelbaum}, {Seljak}, {Guzik},
  {Padmanabhan}, {Blake}, {Brinkmann}, {Bud{\'a}vari}, {Connolly}, {Csabai},
  {Scranton}, \& {Szalay}}]{HS04}
{Hirata}, C.~M., {Mandelbaum}, R., {Seljak}, U., {et~al.} 2004, \mnras, 353,
  529

\bibitem[{{Hoekstra} \& {Jain}(2008)}]{Review_Hoekstra_2008}
{Hoekstra}, H. \& {Jain}, B. 2008, Annual Review of Nuclear and Particle
  Science, 58, 99

\bibitem[{{Ivezic} {et~al.}(2008){Ivezic}, {Tyson}, {Abel}, {Acosta},
  {Allsman}, {AlSayyad}, {Anderson}, {Andrew}, {Angel}, {Angeli}, {Ansari},
  {Antilogus}, {Arndt}, {Astier}, {Aubourg}, {Axelrod}, {Bard}, {Barr},
  {Barrau}, {Bartlett}, {Bauman}, {Beaumont}, {Becker}, {Becla}, {Beldica},
  {Bellavia}, {Blanc}, {Blandford}, {Bloom}, {Bogart}, {Borne}, {Bosch},
  {Boutigny}, {Brandt}, {Brown}, {Bullock}, {Burchat}, {Burke}, {Cagnoli},
  {Calabrese}, {Chandrasekharan}, {Chesley}, {Cheu}, {Chiang}, {Claver},
  {Connolly}, {Cook}, {Cooray}, {Covey}, {Cribbs}, {Cui}, {Cutri}, {Daubard},
  {Daues}, {Delgado}, {Digel}, {Doherty}, {Dubois}, {Dubois-Felsmann},
  {Durech}, {Eracleous}, {Ferguson}, {Frank}, {Freemon}, {Gangler}, {Gawiser},
  {Geary}, {Gee}, {Geha}, {Gibson}, {Gilmore}, {Glanzman}, {Goodenow},
  {Gressler}, {Gris}, {Guyonnet}, {Hascall}, {Haupt}, {Hernandez}, {Hogan},
  {Huang}, {Huffer}, {Innes}, {Jacoby}, {Jain}, {Jee}, {Jernigan},
  {Jevremovic}, {Johns}, {Jones}, {Juramy-Gilles}, {Juric}, {Kahn}, {Kalirai},
  {Kallivayalil}, {Kalmbach}, {Kantor}, {Kasliwal}, {Kessler}, {Kirkby},
  {Knox}, {Kotov}, {Krabbendam}, {Krughoff}, {Kubanek}, {Kuczewski},
  {Kulkarni}, {Lambert}, {Le Guillou}, {Levine}, {Liang}, {Lim}, {Lintott},
  {Lupton}, {Mahabal}, {Marshall}, {Marshall}, {May}, {McKercher}, {Migliore},
  {Miller}, {Mills}, {Monet}, {Moniez}, {Neill}, {Nief}, {Nomerotski},
  {Nordby}, {O'Connor}, {Oliver}, {Olivier}, {Olsen}, {Ortiz}, {Owen}, {Pain},
  {Peterson}, {Petry}, {Pierfederici}, {Pietrowicz}, {Pike}, {Pinto}, {Plante},
  {Plate}, {Price}, {Prouza}, {Radeka}, {Rajagopal}, {Rasmussen}, {Regnault},
  {Ridgway}, {Ritz}, {Rosing}, {Roucelle}, {Rumore}, {Russo}, {Saha},
  {Sassolas}, {Schalk}, {Schindler}, {Schneider}, {Schumacher}, {Sebag},
  {Sembroski}, {Seppala}, {Shipsey}, {Silvestri}, {Smith}, {Smith}, {Strauss},
  {Stubbs}, {Sweeney}, {Szalay}, {Takacs}, {Thaler}, {Van Berg}, {Vanden Berk},
  {Vetter}, {Virieux}, {Xin}, {Walkowicz}, {Walter}, {Wang}, {Warner},
  {Willman}, {Wittman}, {Wolff}, {Wood-Vasey}, {Yoachim}, {Zhan}, \& {for the
  LSST Collaboration}}]{LSST_2008}
{Ivezic}, Z., {Tyson}, J.~A., {Abel}, B., {et~al.} 2008, ArXiv e-prints
  [arXiv:0805.2366v4]

\bibitem[{{Jarvis} {et~al.}(2016){Jarvis}, {Sheldon}, {Zuntz}, {Kacprzak},
  {Bridle}, {Amara}, {Armstrong}, {Becker}, {Bernstein}, {Bonnett}, {Chang},
  {Das}, {Dietrich}, {Drlica-Wagner}, {Eifler}, {Gangkofner}, {Gruen},
  {Hirsch}, {Huff}, {Jain}, {Kent}, {Kirk}, {MacCrann}, {Melchior}, {Plazas},
  {Refregier}, {Rowe}, {Rykoff}, {Samuroff}, {S{\'a}nchez}, {Suchyta},
  {Troxel}, {Vikram}, {Abbott}, {Abdalla}, {Allam}, {Annis}, {Benoit-L{\'e}vy},
  {Bertin}, {Brooks}, {Buckley-Geer}, {Burke}, {Capozzi}, {Carnero Rosell},
  {Carrasco Kind}, {Carretero}, {Castander}, {Clampitt}, {Crocce}, {Cunha},
  {D'Andrea}, {da Costa}, {DePoy}, {Desai}, {Diehl}, {Doel}, {Fausti Neto},
  {Flaugher}, {Fosalba}, {Frieman}, {Gaztanaga}, {Gerdes}, {Gruendl},
  {Gutierrez}, {Honscheid}, {James}, {Kuehn}, {Kuropatkin}, {Lahav}, {Li},
  {Lima}, {March}, {Martini}, {Miquel}, {Mohr}, {Neilsen}, {Nord}, {Ogando},
  {Reil}, {Romer}, {Roodman}, {Sako}, {Sanchez}, {Scarpine}, {Schubnell},
  {Sevilla-Noarbe}, {Smith}, {Soares-Santos}, {Sobreira}, {Swanson}, {Tarle},
  {Thaler}, {Thomas}, {Walker}, \& {Wechsler}}]{DES_2016_cat}
{Jarvis}, M., {Sheldon}, E., {Zuntz}, J., {et~al.} 2016, \mnras, 460, 2245

\bibitem[{{Kacprzak} {et~al.}(2012){Kacprzak}, {Zuntz}, {Rowe}, {Bridle},
  {Refregier}, {Amara}, {Voigt}, \& {Hirsch}}]{Noise_Kacprzak12}
{Kacprzak}, T., {Zuntz}, J., {Rowe}, B., {et~al.} 2012, \mnras, 427, 2711

\bibitem[{{Kaiser} {et~al.}(1995){Kaiser}, {Squires}, \&
  {Broadhurst}}]{KSB_1995}
{Kaiser}, N., {Squires}, G., \& {Broadhurst}, T. 1995, \apj, 449, 460

\bibitem[{{Kitching} {et~al.}(2012){Kitching}, {Balan}, {Bridle}, {Cantale},
  {Courbin}, {Eifler}, {Gentile}, {Gill}, {Harmeling}, {Heymans}, {Hirsch},
  {Honscheid}, {Kacprzak}, {Kirkby}, {Margala}, {Massey}, {Melchior},
  {Nurbaeva}, {Patton}, {Rhodes}, {Rowe}, {Taylor}, {Tewes}, {Viola},
  {Witherick}, {Voigt}, {Young}, \& {Zuntz}}]{GREAT10_2012}
{Kitching}, T.~D., {Balan}, S.~T., {Bridle}, S., {et~al.} 2012, \mnras, 423,
  3163

\bibitem[{{Kitching} {et~al.}(2008){Kitching}, {Miller}, {Heymans}, {van
  Waerbeke}, \& {Heavens}}]{Lensfit_Kitching08}
{Kitching}, T.~D., {Miller}, L., {Heymans}, C.~E., {van Waerbeke}, L., \&
  {Heavens}, A.~F. 2008, \mnras, 390, 149

\bibitem[{{Kuijken}(1999)}]{KK_1999}
{Kuijken}, K. 1999, \aap, 352, 355

\bibitem[{{Kuijken}(2006)}]{KK_2006}
{Kuijken}, K. 2006, \aap, 456, 827

\bibitem[{{Kuijken} {et~al.}(2015){Kuijken}, {Heymans}, {Hildebrandt},
  {Nakajima}, {Erben}, {de Jong}, {Viola}, {Choi}, {Hoekstra}, {Miller}, {van
  Uitert}, {Amon}, {Blake}, {Brouwer}, {Buddendiek}, {Conti}, {Eriksen},
  {Grado}, {Harnois-D{\'e}raps}, {Helmich}, {Herbonnet}, {Irisarri},
  {Kitching}, {Klaes}, {La Barbera}, {Napolitano}, {Radovich}, {Schneider},
  {Sif{\'o}n}, {Sikkema}, {Simon}, {Tudorica}, {Valentijn}, {Verdoes Kleijn},
  \& {van Waerbeke}}]{KIDS_DR2}
{Kuijken}, K., {Heymans}, C., {Hildebrandt}, H., {et~al.} 2015, \mnras, 454,
  3500

\bibitem[{{Lambas} {et~al.}(1992){Lambas}, {Maddox}, \& {Loveday}}]{LML92}
{Lambas}, D.~G., {Maddox}, S.~J., \& {Loveday}, J. 1992, \mnras, 258, 404

\bibitem[{{Laureijs} {et~al.}(2011){Laureijs}, {Amiaux}, {Arduini},
  {Augu{\`e}res}, {Brinchmann}, {Cole}, {Cropper}, {Dabin}, {Duvet}, {Ealet},
  \& et~al.}]{Euclid_2011}
{Laureijs}, R., {Amiaux}, J., {Arduini}, S., {et~al.} 2011, ArXiv e-prints
  [arXiv:1110.3193]

\bibitem[{{Leauthaud} {et~al.}(2007){Leauthaud}, {Massey}, {Kneib}, {Rhodes},
  {Johnston}, {Capak}, {Heymans}, {Ellis}, {Koekemoer}, {Le F{\`e}vre},
  {Mellier}, {R{\'e}fr{\'e}gier}, {Robin}, {Scoville}, {Tasca}, {Taylor}, \&
  {Van Waerbeke}}]{COSMOS_2007}
{Leauthaud}, A., {Massey}, R., {Kneib}, J.-P., {et~al.} 2007, \apjs, 172, 219

\bibitem[{Lee \& Schachter(1980)}]{CHP_tri}
Lee, D.~T. \& Schachter, B.~J. 1980, International Journal of Computer {\&}
  Information Sciences, 9, 219–242

\bibitem[{{Lee}(2007)}]{CHP_Lee}
{Lee}, H. 2007, in Astronomical Society of the Pacific Conference Series, Vol.
  371, Statistical Challenges in Modern Astronomy IV, ed. G.~J. {Babu} \& E.~D.
  {Feigelson}, 425

\bibitem[{{Mandelbaum} {et~al.}(2017){Mandelbaum}, {Miyatake}, {Hamana},
  {Oguri}, {Simet}, {Armstrong}, {Bosch}, {Murata}, {Lanusse}, {Leauthaud},
  {Coupon}, {More}, {Takada}, {Miyazaki}, {Speagle}, {Shirasaki}, {Sif{\'o}n},
  {Huang}, {Nishizawa}, {Medezinski}, {Okura}, {Okabe}, {Czakon}, {Takahashi},
  {Coulton}, {Hikage}, {Komiyama}, {Lupton}, {Strauss}, {Tanaka}, \&
  {Utsumi}}]{HSC_DR1}
{Mandelbaum}, R., {Miyatake}, H., {Hamana}, T., {et~al.} 2017, ArXiv e-prints
  [arXiv:1705.06745]

\bibitem[{{Mandelbaum} {et~al.}(2015){Mandelbaum}, {Rowe}, {Armstrong}, {Bard},
  {Bertin}, {Bosch}, {Boutigny}, {Courbin}, {Dawson}, {Donnarumma}, {Fenech
  Conti}, {Gavazzi}, {Gentile}, {Gill}, {Hogg}, {Huff}, {Jee}, {Kacprzak},
  {Kilbinger}, {Kuntzer}, {Lang}, {Luo}, {March}, {Marshall}, {Meyers},
  {Miller}, {Miyatake}, {Nakajima}, {Ngol{\'e} Mboula}, {Nurbaeva}, {Okura},
  {Paulin-Henriksson}, {Rhodes}, {Schneider}, {Shan}, {Sheldon}, {Simet},
  {Starck}, {Sureau}, {Tewes}, {Zarb Adami}, {Zhang}, \& {Zuntz}}]{GREAT3_2015}
{Mandelbaum}, R., {Rowe}, B., {Armstrong}, R., {et~al.} 2015, \mnras, 450, 2963

\bibitem[{{Massey} {et~al.}(2007){Massey}, {Heymans}, {Berg{\'e}}, {Bernstein},
  {Bridle}, {Clowe}, {Dahle}, {Ellis}, {Erben}, {Hetterscheidt}, {High},
  {Hirata}, {Hoekstra}, {Hudelot}, {Jarvis}, {Johnston}, {Kuijken},
  {Margoniner}, {Mandelbaum}, {Mellier}, {Nakajima}, {Paulin-Henriksson},
  {Peeples}, {Roat}, {Refregier}, {Rhodes}, {Schrabback}, {Schirmer}, {Seljak},
  {Semboloni}, \& {van Waerbeke}}]{STEP2_2007}
{Massey}, R., {Heymans}, C., {Berg{\'e}}, J., {et~al.} 2007, \mnras, 376, 13

\bibitem[{{Melchior} \& {Viola}(2012)}]{Noise_Melchior12}
{Melchior}, P. \& {Viola}, M. 2012, \mnras, 424, 2757

\bibitem[{{Miller} {et~al.}(2013){Miller}, {Heymans}, {Kitching}, {van
  Waerbeke}, {Erben}, {Hildebrandt}, {Hoekstra}, {Mellier}, {Rowe}, {Coupon},
  {Dietrich}, {Fu}, {Harnois-D{\'e}raps}, {Hudson}, {Kilbinger}, {Kuijken},
  {Schrabback}, {Semboloni}, {Vafaei}, \& {Velander}}]{CFHTLenS_Miller13}
{Miller}, L., {Heymans}, C., {Kitching}, T.~D., {et~al.} 2013, \mnras, 429,
  2858

\bibitem[{{Miller} {et~al.}(2007){Miller}, {Kitching}, {Heymans}, {Heavens}, \&
  {van Waerbeke}}]{Lensfit_Miller07}
{Miller}, L., {Kitching}, T.~D., {Heymans}, C., {Heavens}, A.~F., \& {van
  Waerbeke}, L. 2007, \mnras, 382, 315

\bibitem[{Mosteller \& Tukey(1977)}]{BI2_1977}
Mosteller, F. \& Tukey, J.~W. 1977, Data Analysis and Regression: a Second
  Course in Statistics (Reading, MA: Addison Wesley), p. 133

\bibitem[{{Preparata} \& {Shamos}(1985)}]{CHP_alg}
{Preparata}, F.~P. \& {Shamos}, M.~I. 1985, Computational Geometry (New York:
  Springer), 95--149

\bibitem[{Rao(1945)}]{Rao_1945}
Rao, C.~R. 1945, Bull. Calcutta Math. Soc., 37, 81

\bibitem[{{Refregier} \& {Bacon}(2003)}]{Refregier_2003}
{Refregier}, A. \& {Bacon}, D. 2003, \mnras, 338, 48

\bibitem[{{Refregier} {et~al.}(2012){Refregier}, {Kacprzak}, {Amara}, {Bridle},
  \& {Rowe}}]{Noise_Refregier12}
{Refregier}, A., {Kacprzak}, T., {Amara}, A., {Bridle}, S., \& {Rowe}, B. 2012,
  \mnras, 425, 1951

\bibitem[{{Rhodes} {et~al.}(2000){Rhodes}, {Refregier}, \&
  {Groth}}]{Rhodes_2000}
{Rhodes}, J., {Refregier}, A., \& {Groth}, E.~J. 2000, \apj, 536, 79

\bibitem[{{Rodr{\'{\i}}guez} \& {Padilla}(2013)}]{Rodriguez_intsh}
{Rodr{\'{\i}}guez}, S. \& {Padilla}, N.~D. 2013, \mnras, 434, 2153

\bibitem[{{Schneider}(2006)}]{Review_Schneider_2006}
{Schneider}, P. 2006, in Saas-Fee Advanced Course 33: Gravitational Lensing:
  Strong, Weak and Micro, ed. G.~{Meylan}, P.~{Jetzer}, P.~{North},
  P.~{Schneider}, C.~S. {Kochanek}, \& J.~{Wambsganss}, 269--451

\bibitem[{{Seitz} \& {Schneider}(1997)}]{SS3_97}
{Seitz}, C. \& {Schneider}, P. 1997, \aap, 318, 687

\bibitem[{{Stark}(1977)}]{Stark77}
{Stark}, A.~A. 1977, \apj, 213, 368

\bibitem[{{Tyson} {et~al.}(1990){Tyson}, {Valdes}, \& {Wenk}}]{Tyson_1990}
{Tyson}, J.~A., {Valdes}, F., \& {Wenk}, R.~A. 1990, \apjl, 349, L1

\bibitem[{{Van Waerbeke} {et~al.}(2010){Van Waerbeke}, {Hildebrandt}, {Ford},
  \& {Milkeraitis}}]{Mag_Waerbeke_2010}
{Van Waerbeke}, L., {Hildebrandt}, H., {Ford}, J., \& {Milkeraitis}, M. 2010,
  \apjl, 723, L13

\bibitem[{{Van Waerbeke} {et~al.}(2000){Van Waerbeke}, {Mellier}, {Erben},
  {Cuillandre}, {Bernardeau}, {Maoli}, {Bertin}, {McCracken}, {Le F{\`e}vre},
  {Fort}, {Dantel-Fort}, {Jain}, \& {Schneider}}]{Waerbeke_2000}
{Van Waerbeke}, L., {Mellier}, Y., {Erben}, T., {et~al.} 2000, \aap, 358, 30

\bibitem[{{Voigt} \& {Bridle}(2010)}]{Voigt_2010}
{Voigt}, L.~M. \& {Bridle}, S.~L. 2010, \mnras, 404, 458

\bibitem[{{Zhang} {et~al.}(2015){Zhang}, {Luo}, \& {Foucaud}}]{Zhang_2015}
{Zhang}, J., {Luo}, W., \& {Foucaud}, S. 2015, \jcap, 1, 024

\bibitem[{{Zhang} \& {Zhang}(2016)}]{Zhang_2016}
{Zhang}, J. \& {Zhang}, P. 2016, ArXiv e-prints [arXiv:1606.09397]

\end{thebibliography}

\appendix

\section{Bias estimations} \label{app:bias}

In Table \ref{tab:bias} are given the mean-bias and CHP-bias for each estimator, with and without de-shearing. This multiplicative bias $m$ is defined by Equation \ref{eq:bias} as $e_{\mathrm{fit}}=(1+m)e_{\mathrm{in}}$.

\begin{table}[h]
\caption{Results for bias estimations for each estimator, with and without de-shearing. We have used simulated projected ellipsoids as intrinsic ellipticities, with and without added noise. Estimation bias is given in terms of a multiplicative component $m$ as defined in Equation \ref{eq:bias}. Numbers in parentheses reflect the standard uncertainty in the last digit.}
\label{tab:bias}
\centering
\begin{tabular}{rcc}
\hline
\hline
Estim. & $m_{\mu}$  & $m_{\mathrm{CHP}}$  \\
\hline
\multicolumn{3}{c}{\em Simulated ellipticities}  \\
Mean    & -0.0003(4)  & -0.0007(7)  \\
de-$g$  & 0.0006(6)   & 0.0004(4)   \\
        &             &             \\
LAD     & 0.0172(4)   & 0.0154(5)   \\
de-$g$  & -0.0000(2)  & -0.0005(2)  \\
        &             &             \\
CHP     & 0.0097(2)   & 0.0052(5)   \\
de-$g$  & 0.0000(2)   & 0.0001(2)   \\
        &             &             \\
BI      & 0.0426(7)   & 0.0421(9)   \\
de-$g$  & -0.0002(6)  & 0.0002(5)   \\
\hline                                                           
\multicolumn{3}{c}{\em Added noise}  \\
Mean    & -0.0404(7)  & -0.040(1)   \\
de-$g$  & -0.045(2)   & -0.045(2)   \\
        &             &             \\
LAD     & 0.0114(6)   & 0.011(1)    \\
de-$g$  & -0.0308(8)  & -0.031(1)   \\
        &             &             \\
CHP     & 0.0053(5)   & 0.004(1)    \\
de-$g$  & -0.030(1)   & -0.028(2)   \\
        &             &             \\
BI      & 0.0066(8)   & 0.007(1)    \\
de-$g$  & -0.040(1)   & -0.040(2)   \\
\hline                                                           
\end{tabular}                                                   
\end{table}

\section{Efficiency estimations} \label{app:effi}

In Tables \ref{tab:LAD}, \ref{tab:BI}, and \ref{tab:CHP}, we summarize the full results for the relative efficiencies of each estimator. We determine relative efficiencies for coverages of $25\%$, $50\%$, and $75\%$, corresponding to the MAD and the first and third quartiles, and $38.3\%$, $68.3\%$, $86.6\%$, and $95.4\%$, which would correspond to steps of $0.5\sigma$ in case of a Gaussian distribution with variance $\sigma^2$.

For easy reference, we also indicate how much the scale of the estimator distribution would improve, in percentages of the scale of the distribution of the mean estimator,\begin{equation}
\label{imp_scale}
\Delta s_p=\frac{s_{\hat e,p}}{s_\mu}-1 \quad (\mathrm{in }\; \%)
\end{equation}Since a higher efficiency means a smaller scale and therefore a more `trustworthy' estimate, this is an intuitive, albeit rough indication of the change in error bars.

\begin{table*}[h]
\caption{Results for scales of fixed coverage for LAD. For each sample distribution, the relative efficiencies $\eta$ are given first, and the (more intuitive) relative change in estimator distribution scale is given second, in percentages of the distribution scale of the mean. Numbers in parentheses reflect the standard uncertainty in the last digit.}
\label{tab:LAD}
\centering
\begin{tabular}{lccccccc}
\hline
\hline
Distribution & $\eta_{25} $ & $\eta_{38.3} $ & $\eta_{50} $ & $\eta_{68.3} $ & $\eta_{75} $ & $\eta_{86.6} $ & $\eta_{95.4} $ \\
             & $\Delta s_{25}(\% ) $ & $\Delta s_{38.3}(\% ) $ & $\Delta s_{50}(\% ) $ & $\Delta s_{68.3}(\% ) $ & $\Delta s_{75}(\% ) $ & $\Delta s_{86.6}(\% ) $ & $\Delta s_{95.4}(\% ) $ \\
\hline
\multicolumn{8}{c}{\em Simulated ellipticities}  \\
Gaussian    & 0.79(3)  & 0.79(3)  & 0.81(3)  & 0.83(2)  & 0.84(3)  & 0.81(2)  & 0.83(3)  \\ 
            & +12(2)   & +12(2)   & +11(2)   & +9(1)    & +9(2)    & +11(1)   & +10(2)   \\ 
Uniform $q$ & 7.2(4)   & 6.1(3)   & 5.4(1)   & 4.55(9)  & 4.3(1)   & 3.69(9)  & 3.22(6)  \\
            & -63(1)   & -59.4(8) & -56.9(5) & -53.1(5) & -51.7(7) & -48.0(6) & -44.3(5) \\
Elliptical  & 2.7(1)   & 2.5(1)   & 2.4(1)   & 2.26(8)  & 2.20(9)  & 2.07(8)  & 1.88(7)  \\ 
            & -39(2)   & -36(2)   & -35(1)   & -34(1)   & -33(1)   & -31(1)   & -27(1)   \\ 
Disk        & 6.3(3)   & 5.8(2)   & 5.5(2)   & 5.1(2)   & 4.9(1)   & 4.7(1)   & 4.3(1)   \\ 
            & -60.1(9) & -58.5(8) & -57.3(6) & -55.6(7) & -54.7(7) & -53.9(7) & -52.0(7) \\ 
Combined    & 5.4(2)   & 5.0(3)   & 4.9(2)   & 4.6(2)   & 4.4(1)   & 4.1(1)   & 4.0(2)   \\ 
            & -57.2(8) & -55(1)   & -55(1)   & -53.6(8) & -52.2(7) & -50.8(7) & -50(1)   \\ 
\hline                                                           
\multicolumn{8}{c}{\em Added noise}  \\
Uniform $q$ & 1.37(7)  & 1.38(5)  & 1.37(5)  & 1.38(3)  & 1.37(3)  & 1.33(3)  & 1.29(4)  \\
            & -15(2)   & -15(2)   & -14(2)   & -15(1)   & -14(1)   & -13(1)   & -12(1)   \\
Elliptical  & 1.19(6)  & 1.18(5)  & 1.20(6)  & 1.17(3)  & 1.18(4)  & 1.19(3)  & 1.16(4)  \\ 
            & -8(2)    & -8(2)    & -9(2)    & -8(1)    & -8(1)    & -8(1)    & -7(2)    \\ 
Disk        & 1.48(8)  & 1.49(7)  & 1.54(7)  & 1.55(6)  & 1.57(5)  & 1.59(5)  & 1.52(4)  \\ 
            & -18(2)   & -18(2)   & -19(2)   & -20(1)   & -20(1)   & -21(1)   & -19(1)   \\ 
Combined    & 1.55(5)  & 1.53(6)  & 1.51(8)  & 1.53(5)  & 1.51(5)  & 1.48(6)  & 1.46(5)  \\ 
            & -20(1)   & -19(2)   & -19(2)   & -19(1)   & -19(1)   & -18(20   & -17(1)   \\ 
\hline                                                           
\multicolumn{8}{c}{\em Full CFHTLenS data}  \\
All         & 1.17(6)  & 1.19(4)  & 1.19(3)  & 1.21(3)  & 1.20(2)  & 1.20(3)  & 1.12(3)  \\ 
            & -8(3)    & -8(1)    & -8(1)    & -9(1)    & -8.6(9)  & -9(1)    & -6(1)    \\ 
Red         & 1.32(7)  & 1.33(6)  & 1.35(5)  & 1.26(4)  & 1.24(5)  & 1.19(5)  & 1.19(8)  \\ 
            & -13(2)   & -13(2)   & -14(1)   & -11(1)   & -10(2)   & -8(2)    & -8(3)    \\ 
Blue        & 1.28(4)  & 1.24(5)  & 1.24(4)  & 1.20(3)  & 1.18(3)  & 1.19(5)  & 1.19(8)  \\ 
            & -12(1)   & -10(2)   & -10(1)   & -9(1)    & -8(1)    & -8(2)    & -8(3)    \\ 
Sbc         & 1.23(5)  & 1.19(3)  & 1.19(5)  & 1.18(2)  & 1.17(3)  & 1.18(3)  & 1.13(6)  \\ 
            & -10(2)   & -8(1)    & -9(2)    & -7.8(6)  & -7.4(9)  & -8(1)    & -6(2)    \\ 
Scd         & 1.21(4)  & 1.20(4)  & 1.21(4)  & 1.18(4)  & 1.16(3)  & 1.19(6)  & 1.12(7)  \\ 
            & -9(2)    & -9(2)    & -9(2)    & -8(1)    & -7(1)    & -8(2)    & -6(3)    \\ 
\hline                                                           
\multicolumn{8}{c}{\em Conservative selection of CFHTLenS data}  \\
All         & 1.82(7)  & 1.77(6)  & 1.72(4)  & 1.67(5)  & 1.70(3)  & 1.65(2)  & 1.67(6)  \\ 
            & -26(1)   & -25(1)   & -24(1)   & -23(1)   & -23.3(7) & -22.2(4) & -23(1)   \\ 
Red         & 1.8(1)   & 1.68(9)  & 1.66(7)  & 1.64(6)  & 1.64(6)  & 1.66(6)  & 1.58(6)  \\ 
            & -26(3)   & -23(2)   & -22(2)   & -22(1)   & -22(1)   & -22(1)   & -21(2)   \\ 
Blue        & 1.73(4)  & 1.63(5)  & 1.62(5)  & 1.65(4)  & 1.64(6)  & 1.64(5)  & 1.61(9)  \\ 
            & -24.0(8) & -22(1)   & -21(1)   & -22(1)   & -22(1)   & -22(1)   & -21(2)   \\ 
Sbc         & 1.63(8)  & 1.6(1)   & 1.59(6)  & 1.55(4)  & 1.52(4)  & 1.49(4)  & 1.47(7)  \\ 
            & -22(2)   & -22(2)   & -21(2)   & -20(1)   & -19(1)   & -18(1)   & -18(2)   \\ 
Scd         & 1.8(1)   & 1.75(9)  & 1.70(6)  & 1.67(4)  & 1.6(4)   & 1.65(3)  & 1.55(5)  \\
            & -25(3)   & -24(2)   & -23(1)   & -22.6(9) & -21.9(9) & -22.1(7) & -20(1)   \\
\hline                                                           
\end{tabular}                                                    
\end{table*}

\begin{table*}[h]
\caption{Results for scales of fixed coverage for the biweight. For each sample distribution, the relative efficiencies $\eta$ are given first, and the (more intuitive) relative change in estimator distribution scale is given second, in percentages of the distribution scale of the mean. Numbers in parentheses reflect the standard uncertainty in the last digit.}
\label{tab:BI}
\centering
\begin{tabular}{lccccccc}
\hline
\hline
Distribution & $\eta_{25} $ & $\eta_{38.3} $ & $\eta_{50} $ & $\eta_{68.3} $ & $\eta_{75} $ & $\eta_{86.6} $ & $\eta_{95.4} $ \\
             & $\Delta s_{25}(\% ) $ & $\Delta s_{38.3}(\% ) $ & $\Delta s_{50}(\% ) $ & $\Delta s_{68.3}(\% ) $ & $\Delta s_{75}(\% ) $ & $\Delta s_{86.6}(\% ) $ & $\Delta s_{95.4}(\% ) $ \\
\hline
\multicolumn{8}{c}{\em Simulated ellipticities}  \\
Gaussian    & 0.99(6) & 1.02(6) & 1.00(4) & 1.03(4) & 1.02(5) & 1.03(4) & 1.05(4) \\ 
            & +0(3)   & -1(3)   & -0(2)   & -1(2)   & -1(2)   & -1(2)   & -2(2)   \\ 
Uniform $q$ & 1.2(1)  & 1.16(5) & 1.19(2) & 1.14(3) & 1.12(2) & 1.14(5) & 1.17(6) \\
            & -10(3)  & -7(2)   & -8.2(8) & -6(1)   & -6(1)   & -6(2)   & -7(2)   \\
Elliptical  & 1.4(2)  & 1.3(1)  & 1.31(8) & 1.31(8) & 1.30(6) & 1.34(9) & 1.32(3) \\ 
            & -14(5)  & -13(2)  & -13(3)  & -13(2)  & -12(2)  & -14(3)  & -13(1)  \\ 
Disk        & 1.6(1)  & 1.60(7) & 1.52(6) & 1.49(6) & 1.49(6) & 1.49(5) & 1.46(4) \\ 
            & -20(3)  & -21(2)  & -19(2)  & -18(2)  & -18(2)  & -18(1)  & -17(1)  \\ 
Combined    & 1.54(8) & 1.5(1)  & 1.5(1)  & 1.46(5) & 1.47(6) & 1.47(5) & 1.45(8) \\ 
            & -19(2)  & -19(3)  & -18(3)  & -17(2)  & -17(2)  & -17(1)  & -17(2)  \\ 
\hline                                                           
\multicolumn{8}{c}{\em Added noise}  \\
Uniform $q$ & 1.09(6) & 1.11(5) & 1.11(7) & 1.09(4) & 1.09(5) & 1.09(6) & 1.10(4) \\
            & -4(2)   & -5(2)   & -5(3)   & -4(2)   & -4(2)   & -4(2)   & -5(2)   \\
Elliptical  & 1.23(8) & 1.26(9) & 1.23(9) & 1.248)  & 1.23(8) & 1.24(6) & 1.22(4) \\ 
            & -10(3)  & -11(3)  & -10(3)  & -10(3)  & -10(3)  & -10(2)  & -9(2)   \\ 
Disk        & 1.25(6) & 1.24(9) & 1.24(7) & 1.23(4) & 1.23(4) & 1.23(6) & 1.19(4) \\ 
            & -10(2)  & -10(3)  & -10(2)  & -10(1)  & -10(2)  & -10(2)  & -8(1)   \\ 
Combined    & 1.23(7) & 1.25(6) & 1.24(7) & 1.26(6) & 1.26(7) & 1.2397) & 1.22(6) \\ 
            & -10(2)  & -10(2)  & -10(3)  & -11(2)  & -11(2)  & -10(3)  & -9(2)   \\ 
\hline                                                           
\multicolumn{8}{c}{\em Full CFHTLenS data}  \\
All         & 1.06(7) & 1.06(4) & 1.05(3) & 1.10(3) & 1.09(3) & 1.08(3) & 1.06(2) \\ 
            & -3(3)   & -3(2)   & -2(2)   & -5(1)   & -4(2)   & -4(1)   & -3.1(8) \\ 
Red         & 1.1(1)  & 1.10(8) & 1.09(5) & 1.08(5) & 1.08(6) & 1.08(6) & 1.1(1)  \\ 
            & -5(6)   & -5(3)   & -4(2)   & -4(2)   & -4(3)   & -4(3)   & -4(5)   \\ 
Blue        & 1.09(3) & 1.06(4) & 1.05(7) & 1.08(5) & 1.10(5) & 1.10(7) & 1.07(8) \\ 
            & -4(1)   & -3(2)   & -3(3)   & -4(2)   & -5(2)   & -5(3)   & -3(4)   \\ 
Sbc         & 1.12(9) & 1.08(7) & 1.07(6) & 1.06(3) & 1.07(4) & 1.06(4) & 1.03(5) \\ 
            & -5(4)   & -4(3)   & -3(3)   & -3(1)   & -3(2)   & -3(2)   & -2(2)   \\ 
Scd         & 1.02(4) & 1.03(4) & 1.05(4) & 1.05(4) & 1.04(5) & 1.06(8) & 1.02(9) \\ 
            & -1(2)   & -1(2)   & -3(2)   & -2(2)   & -2(2)   & -3(4)   & -1(4)   \\ 
\hline                                                           
\multicolumn{8}{c}{\em Conservative selection of CFHTLenS data}  \\
All         & 1.22(6) & 1.23(3) & 1.21(4) & 1.19(5) & 1.22(5) & 1.22(3) & 1.21(7) \\ 
            & -10(2)  & -10(1)  & -9(2)   & -8(2)   & -9(2)   & -9(1)   & -9(3)   \\ 
Red         & 1.12(5) & 1.10(5) & 1.12(5) & 1.11(4) & 1.11(4) & 1.15(6) & 1.14(50 \\ 
            & -6(2)   & -5(2)   & -5(2)   & -5(2)   & -5(2)   & -7(2)   & -6(2)   \\ 
Blue        & 1.27(8) & 1.22(7) & 1.23(8) & 1.23(7) & 1.23(5) & 1.23(5) & 1.21(6) \\ 
            & -11(3)  & -9(3)   & -10(3)  & -10(3)  & -10(2)  & -10(2)  & -9(2)   \\ 
Sbc         & 1.14(6) & 1.2(1)  & 1.14(7) & 1.13(3) & 1.13(4) & 1.15(7) & 1.15(4) \\ 
            & -6(2)   & -7(5)   & -6(3)   & -6(1)   & -6(2)   & -7(3)   & -7(1)   \\ 
Scd         & 1.3(1)  & 1.3(1)  & 1.30(9) & 1.28(8) & 1.25(8) & 1.26(6) & 1.28(4) \\
            & -14(4)  & -14(4)  & -12(3)  & -12(3)  & -10(3)  & -11(2)  & -11(1)  \\
\hline                                                           
\end{tabular}                                                    
\end{table*}

\begin{table*}[h]
\caption{Results for scales of fixed coverage for CHP. For each sample distribution, the relative efficiencies $\eta$ are given first, and the (more intuitive) relative change in estimator distribution scale is given second, in percentages of the distribution scale of the mean. Numbers in parentheses reflect the standard uncertainty in the last digit.}
\label{tab:CHP}
\centering
\begin{tabular}{lccccccc}
\hline
\hline
Distribution & $\eta_{25} $ & $\eta_{38.3} $ & $\eta_{50} $ & $\eta_{68.3} $ & $\eta_{75} $ & $\eta_{86.6} $ & $\eta_{95.4} $ \\
             & $\Delta s_{25}(\% ) $ & $\Delta s_{38.3}(\% ) $ & $\Delta s_{50}(\% ) $ & $\Delta s_{68.3}(\% ) $ & $\Delta s_{75}(\% ) $ & $\Delta s_{86.6}(\% ) $ & $\Delta s_{95.4}(\% ) $ \\
\hline
\multicolumn{8}{c}{\em Simulated ellipticities}  \\
Gaussian    & 0.38(3) & 0.38(1)  & 0.38(1)  & 0.39(2)  & 0.38(1)  & 0.37(1) & 0.38(2) \\ 
            & +62(6)  & +62(3)   & +63(3)   & +61(3)   & +63(2)   & +63(2)  & +63(5)  \\ 
Uniform $q$ & 15(2)   & 11.1(9)  & 8.9(5)   & 6.5(5)   & 5.6(4)   & 4.4(3)  & 3.3(3)  \\
            & -74(2)  & -70(1)   & -66.6(9) & -61(1)   & -58(1)   & -53(1)  & -45(2)  \\
Elliptical  & 2.3(2)  & 2.03(8)  & 1.80(7)  & 1.58(5)  & 1.51(6)  & 1.31(8) & 1.13(8) \\ 
            & -33(2)  & -30(1)   & -25(1)   & -21(1)   & -19(2)   & -13(3)  & -6(3)   \\ 
Disk        & 6.8(7)  & 6.0(5)   & 5.4(2)   & 4.6(1)   & 4.3(1)   & 3.8(2)  & 3.2(2)  \\ 
            & -62(2)  & -59(2)   & -57.1(9) & -53.5(6) & -51.6(8) & -49(1)  & -44(2)  \\ 
Combined    & 5.8(3)  & 5.0(1)   & 4.7(2)   & 4.1(1)   & 3.8(1)   & 3.4(1)  & 2.7(2)  \\ 
            & -59(1)  & -55.4(6) & -53.8(8) & -50.4(7) & -48.8(9) & -45(1)  & -39(2)  \\ 
\hline                                                           
\multicolumn{8}{c}{\em Added noise}  \\
Uniform $q$ & 0.85(4) & 0.84(3)  & 0.86(4)  & 0.85(4)  & 0.84(3)  & 0.80(3) & 0.77(40 \\
            & +7(2)   & +9(2)    & +8(3)    & +8(2)    & +9(2)    & +12(2)  & +14(3)  \\
Elliptical  & 0.65(5) & 0.65(6)  & 0.64(4)  & 0.66(3)  & 0.64(3)  & 0.64(3) & 0.64(4) \\ 
            & +24(5)  & +24(6)   & +25(4)   & +24(3)   & +25(3)   & +25(3)  & +25(4)  \\ 
Disk        & 0.88(7) & 0.89(6)  & 0.88(4)  & 0.90(3)  & 0.90(3)  & 0.88(4) & 0.87(6) \\ 
            & +7(4)   & +6(4)    & +6(3)    & +6(2)    & +5(1)    & +6(2)   & +7(4)   \\ 
Combined    & 0.91(6) & 0.92(6)  & 0.89(5)  & 0.91(3)  & 0.89(4)  & 0.88(5) & 0.86(4) \\ 
            & +5(3)   & +4(4)    & +6(3)    & +5(2)    & +6(2)    & +6(3)   & +8(2)   \\ 
\hline                                                           
\multicolumn{8}{c}{\em Full CFHTLenS data}  \\
All         & 0.75(4) & 0.76(3)  & 0.77(2)  & 0.76(2)  & 0.77(2)  & 0.75(3) & 0.71(2) \\ 
            & +16(3)  & +14(2)   & +14(2)   & +15(1)   & +14(1)   & +16(2)  & +19(1)  \\ 
Red         & 0.81(7) & 0.84(7)  & 0.84(4)  & 0.77(4)  & 0.76(5)  & 0.72(4) & 0.72(4) \\ 
            & +11(5)  & +9(4)    & +9(2)    & +14(3)   & +15(4)   & +18(3)  & +18(3)  \\ 
Blue        & 0.81(3) & 0.79(3)  & 0.77(3)  & 0.75(4)  & 0.76(3)  & 0.73(3) & 0.68(6) \\ 
            & +11(2)  & +13(2)   & +14(2)   & +15(3)   & +14(3)   & +17(2)  & +21(5)  \\ 
Sbc         & 0.80(4) & 0.78(3)  & 0.77(3)  & 0.76(3)  & 0.74(2)  & 0.71(3) & 0.68(4) \\ 
            & +12(3)  & +14(2)   & +14(2)   & +15(2)   & +16(2)   & +18(2)  & +21(4)  \\ 
Scd         & 0.79(5) & 0.75(3)  & 0.74(2)  & 0.73(3)  & 0.72(4)  & 0.72(3) & 0.67(5) \\ 
            & +12(3)  & +15(2)   & +16(2)   & +17(2)   & +18(3)   & +18(3)  & +22(5)  \\ 
\hline                                                           
\multicolumn{8}{c}{\em Conservative selection of CFHTLenS data}  \\
All         & 1.18(5) & 1.14(4)  & 1.08(5)  & 1.00(5)  & 0.99(4)  & 0.96(4) & 0.89(5) \\ 
            & -8(2)   & -6(1)    & -4(2)    & 0(3)     & 0(2)     & +2(2)   & +6(3)   \\ 
Red         & 1.3(1)  & 1.21(8)  & 1.15(5)  & 1.05(5)  & 1.02(5)  & 0.99(5) & 0.89(8) \\ 
            & -13(3)  & -9(3)    & -7(2)    & -2(2)    & -1(2)    & 0(2)    & +6(5)   \\ 
Blue        & 1.09(5) & 1.02(6)  & 1.03(7)  & 0.99(5)  & 0.96(5)  & 0.97(5) & 0.93(9) \\ 
            & -4(2)   & -1(3)    & -1(3)    & +1(3)    & +2(3)    & +2(3)   & +4(5)   \\ 
Sbc         & 1.1(1)  & 1.1(1)   & 1.0(1)   & 0.98(5)  & 0.95(5)  & 0.91(4) & 0.88(4) \\ 
            & -4(5)   & -3(5)    & -1(5)    & +1(2)    & +3(3)    & +5(2)   & +7(2)   \\ 
Scd         & 1.11(9) & 1.11(4)  & 1.06(6)  & 1.00(6)  & 0.97(5)  & 0.94(5) & 0.85(6) \\
            & -5(4)   & -5(2)    & -3(3)    & 0(3)     & +1(3)    & +3(2)   & +8(3)   \\
\hline                                                           
\end{tabular}                                                    
\end{table*}

\begin{table*}[h]
\caption{Results for $\eta_{68.3}$ for different sample sizes. Numbers in parentheses reflect the standard uncertainty in the last digit.}
\label{tab:eff_sample}
\centering
\begin{tabular}{llcccccc}
\hline
\hline
Distribution & Estimator & $N=10$  & $N=22$  & $N=46$  & $N=100$ & $N=215$ & $N=464$ \\
\hline
\multicolumn{8}{c}{\em Simulated ellipticities}  \\
Gaussian     & LAD       & 0.80(3) & 0.76(2) & 0.80(3) & 0.83(2) & 0.79(3) & 0.78(3) \\
             & BI        & 1.01(3) & 1.06(3) & 1.03(2) & 1.03(4) & 1.03(5) & 1.07(5) \\
             & CHP       & 0.71(2) & 0.50(1) & 0.44(1) & 0.39(2) & 0.33(2) & 0.29(1) \\
Uniform $q$  & LAD       & 1.84(6) & 2.78(5) & 3.6(1)  & 4.55(9) & 5.8(2)  & 6.6(4)  \\
             & BI        & 1.16(9) & 1.14(4) & 1.20(3) & 1.14(3) & 1.17(4) & 1.18(9) \\
             & CHP       & 1.30(5) & 2.19(7) & 3.6(2)  & 6.5(5)  & 12.3(5) & 22(2)   \\
Combined     & LAD       & 2.94(7) & 3.5(2)  & 4.2(1)  & 4.6(2)  & 4.9(1)  & 5.1(2)  \\
             & BI        & 1.47(7) & 1.46(7) & 1.45(4) & 1.46(5) & 1.56(6) & 1.53(8) \\
             & CHP       & 2.03(9) & 2.8(2)  & 3.6(2)  & 4.1(1)  & 5.1(2)  & 6.0(4)  \\
\hline                                                           
\multicolumn{8}{c}{\em Added noise}  \\
Uniform $q$  & LAD       & 1.08(3) & 1.27(7) & 1.32(6) & 1.38(3) & 1.41(4) & 1.33(4) \\
             & BI        & 1.09(3) & 1.08(7) & 1.15(4) & 1.09(4) & 1.08(4) & 1.14(8) \\
             & CHP       & 0.9(8)  & 0.93(6) & 0.93(3) & 0.85(4) & 0.74(4) & 0.66(4) \\
Combined     & LAD       & 1.28(5) & 1.42(3) & 1.53(5) & 1.53(5) & 1.51(6) & 1.56(5) \\ 
             & BI        & 1.24(4) & 1.23(5) & 1.26(3) & 1.26(6) & 1.25(7) & 1.24(8) \\
             & CHP       & 0.98(9) & 1.04(4) & 0.98(8) & 0.91(3) & 0.78(4) & 0.70(4) \\
\hline                                                             
\multicolumn{8}{c}{\em CFHTLenS data}  \\
All          & LAD       & 0.91(2) & 1.11(5) & 1.16(4) & 1.21(3) & 1.15(3) & 1.21(5) \\ 
             & BI        & 1.05(2) & 1.07(4) & 1.10(6) & 1.10(3) & 1.08(3) & 1.09(8) \\
             & CHP       & 0.77(4) & 0.83(6) & 0.82(4) & 0.76(2) & 0.66(2) & 0.65(4) \\
Subset       & LAD       & 1.39(4) & 1.59(6) & 1.59(5) & 1.67(5) & 1.74(4) & 1.80(4) \\ 
             & BI        & 1.17(3) & 1.18(6) & 1.22(5) & 1.19(5) & 1.21(6) & 1.19(3) \\
             & CHP       & 1.07(4) & 1.12(6) & 1.04(3) & 1.00(5) & 0.99(4) & 0.98(6) \\
\hline                                                           
\end{tabular}                                                   
\end{table*}

\section{Estimations from Fourier mode fitting} \label{app:biasFMF}

In Table \ref{tab:bias_FMF} are given the mean-bias and efficiencies of the LSQ estimator, per mode and per individual amplitude, LAD estimator with and without de-shearing, per mode and per individual amplitude, and the CHP estimator, per individual amplitude. The mean-bias is again given in terms of a multiplicative component $m$ as defined in Equation \ref{eq:bias_FMF}.

\begin{table}[h]
\caption{Results for amplitude estimations for FMF, using different estimators (LSQ, LAD, CHP) and models (per mode or per amplitude). Estimation bias is given in terms of a multiplicative component $m$ as defined in Equation \ref{eq:bias_FMF}. Efficiencies are determined relative to LSQ per individual amplitude. Numbers in parentheses reflect the standard uncertainty in the last digit.}
\label{tab:bias_FMF}
\centering
\begin{tabular}{rrcc}
\hline
\hline
\multicolumn{2}{l}{Estimator} & $m$ & $\eta_{68.3}$ \\
\hline
\multicolumn{4}{c}{\em Simulated ellipticities}  \\
LSQ &        per mode & 0.001(1)  & 1.14(6) \\
    &       amplitude & 0.000(1)  & N.A. \\
LAD &        per mode & 0.0166(6) & 4.6(3) \\
    &       amplitude & 0.046(1)  & 1.31(4) \\
    & de-$g$ per mode & 0.0006(6) & 5.4(4) \\
    &       amplitude & 0.022(1)  & 1.61(5) \\
CHP &   per amplitude & 0.043(2)  & 0.83(5) \\
\hline                                                           
\multicolumn{4}{c}{\em Added noise}  \\
LSQ &        per mode & -0.024(1) & 1.09(7) \\
    &       amplitude & -0.023(2) & N.A. \\
LAD &        per mode & 0.014(1)  & 1.47(9) \\
    &       amplitude & 0.032(2)  & 0.98(7) \\
    & de-$g$ per mode & -0.018(3) & 1.55(9) \\
    &       amplitude & -0.007(2) & 1.14(8) \\
CHP &   per amplitude & 0.015(2)  & 0.41(3) \\
\hline                                                           
\end{tabular}                                                   
\end{table}

\end{document}